\documentclass{article}
\usepackage{fullpage}

\usepackage{hyperref}
\hypersetup{
  colorlinks=true,
  linkcolor=blue,
  citecolor=blue,
  urlcolor=blue,
  filecolor=blue
}

\usepackage{tabularx}
\usepackage{array}
\usepackage{listings}
\usepackage{amsmath}
\usepackage[capitalise,noabbrev]{cleveref}
\usepackage{booktabs}
\usepackage{enumitem}
\usepackage{lipsum}
\usepackage{microtype}
\usepackage{array}
\usepackage{ragged2e} 
\usepackage{csquotes}
\usepackage{longtable}
\usepackage{pdflscape}
\usepackage[table]{xcolor}
\usepackage[
  backend=biber,
  style=alphabetic,
  sorting=nty,        
]{biblatex}
\usepackage{titlesec}
\titlespacing*{\paragraph}{0pt}{1ex plus 0.3ex minus 0.2ex}{0.5em}

\usepackage{caption}
\usepackage{tikz}
\usepackage{pgfplots,pgfplotstable}
\usetikzlibrary{patterns}
\usetikzlibrary{arrows.meta,positioning}
\usetikzlibrary{calc}
\usetikzlibrary{external}
\usepgfplotslibrary{groupplots}
\pgfplotsset{compat=1.18}
\usepackage[T1]{fontenc}

\usepackage{xparse}
\usepackage{xcolor}

\newcommand{\code}[1]{\texttt{#1}}

\newcommand{\Description}[1]{}

\crefname{lstlisting}{listing}{listings}
\Crefname{lstlisting}{Listing}{Listings}

\title{greCAPTCHA: Assessing Understanding as Evidence of Research Authorship Under Generative AI}
\author{
    \large
    Justin Payan$^{1,*}$, 
    B\'alint Gyevn\'ar$^{2,*}$,
    Atoosa Kasirzadeh$^{3}$, 
    Nihar B. Shah$^{1,4}$ \\[0.5em] 
    {\small $^1$~Machine Learning Department, Carnegie Mellon University}\\
    {\small $^2$~Institute for Complex Social Dynamics, Carnegie Mellon University}\\
    {\small$^3$~Departments of Philosophy \& Software and Societal Systems, Carnegie Mellon University}\\     
    {\small$^4$~Computer Science Department, Carnegie Mellon University}\\
    {\small$^*$~Equal contribution} \\[0.5em]
    {\small\texttt{\{jpayan,bgyevnar,akasirza,nihars\}@andrew.cmu.edu}}
}
\date{}

\begin{document}

\newcolumntype{P}[1]{>{\RaggedRight\arraybackslash\hspace{0pt}}p{#1}}
\newcommand{\pquote}[2]{\textit{``#1''} (#2)}
\newcommand{\simplequote}[1]{\textit{``#1''}}

\maketitle

\begin{abstract}  
    \noindent
    Conferences, journals, funders, schools, and universities are struggling with a surge of potentially AI-generated submissions from ostensibly human authors, who may not have exercised sufficient human oversight for their manuscripts. In turn, institutions evaluating submissions can no longer reliably credit expertise based solely on authors' names on submitted work. To address this problem, we propose greCAPTCHA, a proctored assessment approach that measures authors' understanding of research manuscripts via the construct of capacity to verify, which we define as the knowledge and reasoning required to critically assess the contents underlying one's contributions to a manuscript. greCAPTCHA generates questions assessing multiple levels of understanding and provides an evaluative report based on authors' responses. Using a prototype implementation, we conduct a user study and semi-structured interviews with $31$ researchers to evaluate greCAPTCHA. Its automated scores predict which papers were or were not authored by study participants with an AUC of $0.90$. Participants reported positive overall experiences with the system and remarked on the appropriate construct validity for author understanding, while also suggesting important changes to be made before deployment. Our results provide initial evidence that greCAPTCHA can assess manuscript-specific understanding under proctored conditions.
\end{abstract}

\begin{figure}
  \centering
  \input{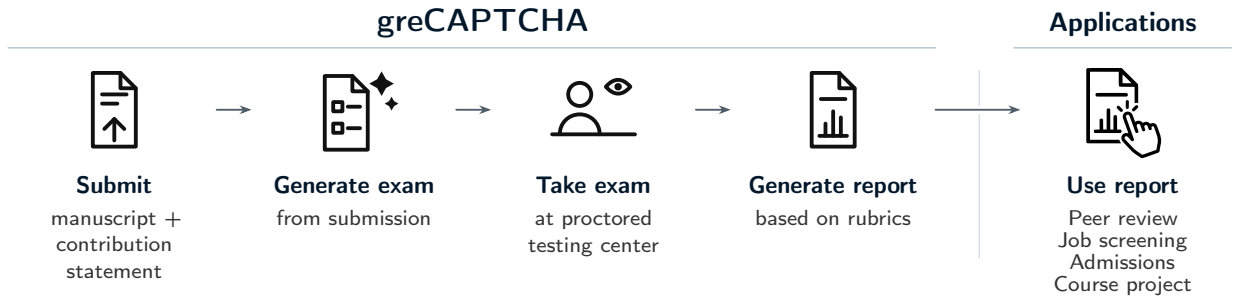}
  \vspace{-1em}
  \caption{The process of assessing author understanding with greCAPTCHA. An author uploads a research manuscript and specifies their specific contributions, from which an exam is generated automatically. The author takes the exam in a proctored setting, either in a testing center (in a university, at a conference, or a dedicated center) or at home~\cite{DuolingoEnglishTest,TOEFLIBTTesting}. The results are then compiled into a report, which can be used for editorial decisions, applicant screening in academic hiring, student admissions assessment, grant proposal screening, or course project verification.}
  \label{fig:greCAPTCHA-teaser}
\end{figure}


\section{Introduction}
\label{sec:introduction}

While the definitions of research authorship are nuanced, they usually require an author to provide original content, substantial contributions, and assume accountability for a new piece of work~\cite{ACMPolicyAuthorship,ICMJERecommendationsDefining}. 
Generative Artificial Intelligence (GenAI) makes it cheap to produce manuscripts that look polished without requiring the author to understand or verify their contents. This shift is exacerbating an existing gap between the quality of submitted manuscripts and the substantive knowledge of their authors.
This gap is present wherever institutions primarily rely on submitted work to assess expertise or award credit, including in scientific publishing, academic hiring, grant evaluation, and education. 
While GenAI can assist in producing a manuscript, we think that for now responsibility for its claims must remain with the human authors.

The incentives and the expanding scale of academic submissions render this accountability gap particularly urgent. 
Acceptance at or rejection from prestigious venues can bring substantial career impacts, and submitting additional manuscripts is almost free.
GenAI can further reduce the cost of creating these submissions, exacerbating the misalignment of incentives to prioritize quantity over quality.
Meanwhile, rapid submission growth enabled by GenAI across research venues places increasing demands on human reviewers~\cite{arxivMonthlySubmissions2026,iclr2026programchairsRetrospectiveICLR20262026,tmlrAnnualAuthorSubmission2026}. 
In education, similar concerns challenge the validity of written assessments and coursework projects that would serve as evidence of students' learning~\cite{chirikovGenerativeAIUse2026}. 
These considerations motivate a pressing question:  
\begin{quote} \emph{How can institutions assess whether the people submitting a manuscript understand their contributions well enough to evaluate and take responsibility for them?}\end{quote}

\subsection{Assessment for Research Authorship}
\label{ssec:intro:exams}

There are various existing approaches to ensuring research integrity. 
Responses include restrictions on submission practices~\cite{communicationchairs2026AIGeneratedPapersNeurIPS2026}, sanctions for policy violations~\cite{dietterichAttentionArxivAuthors2026}, and methods for detecting machine-generated content~\cite{raoDetectingLLMgeneratedPeer2025}. 
These interventions address different aspects of research integrity, but do not directly establish whether authors understand and can evaluate their contributions.
In this paper, we propose a complementary approach based on authors' responsibility to verify the work for which they claim credit.

Drawing on the concept of \emph{evaluative judgment}---the ability to judge the quality of work of oneself and others~\cite{taiDevelopingEvaluativeJudgement2018}---we propose \emph{capacity to verify} as an assessment construct.
We define this capacity as the knowledge and reasoning required to critically assess the claims, methods, and evidence underlying one's specific contributions to a manuscript.
We operationalize it through two dimensions: \emph{comprehension}, that is, understanding the relevant content and its basis; and \emph{justification}, that is, explaining and evaluating the choices underlying the work. 
This capacity is a necessary prerequisite for verification, although demonstrating it does not establish that verification occurred or that the work is correct.
This construct motivates our vision for a new kind of assessment:
\begin{description}
    \item[\textbf{greCAPTCHA.}] Authors demonstrate capacity to verify for their substantial contributions to a manuscript through an assessment tailored to the work and their contributions to it.
\end{description}

Borrowing its name from the Graduate Record Examinations (GRE) used in university admissions, we envision a greCAPTCHA as a proctored exam that asks authors to explain and critically evaluate relevant aspects of their manuscripts. Authors could complete proctored greCAPTCHAs in monitored testing centers at universities, at professional events such as conferences, or through certified at-home assessments \cite{TOEFLIBTTesting,DuolingoEnglishTest}.
Its name also alludes to CAPTCHAs or reCAPTCHAs used to distinguish human users from automated bots. 
However, unlike conventional CAPTCHAs, our objective is to distinguish authors who present AI-generated research as their own with no human contribution or oversight from those who contributed to or at least provided oversight of the research. 

In greCAPTCHA, questions are tailored to authors' contributions, recognizing the division of expertise in collaborative work. 
Grounded in educational assessment~\cite{dawsonDefendingAssessmentSecurity2020}, our approach aims to elicit evidence of understanding while accommodating legitimate uses of GenAI, including language assistance~\cite{gartenbergMoreBetterArtificial2026}, accessibility~\cite{chheda-kotharyQueryingMultimodalScientific2026}, and exploration~\cite{asaiSynthesizingScientificLiterature2026}. 
Much like how students prepare for the GREs by learning, a greCAPTCHA could encourage authors to study their work more carefully, thereby connecting accountability with a broader, positive goal: one of assurance of learning~\cite{dawsonValidityMattersMore2024}. 
Although our experiment initially investigates scientific manuscripts, our proposal is relevant to educational and professional settings in which people submit work for which they are expected to take responsibility---such as academic hiring, student admissions, grant applications, and research proposals. 

Whether greCAPTCHA offers valid indicators of actual human authorship remains an empirical question.
Responses may reflect manuscript-specific understanding, but also domain expertise, question ambiguity, assessment conditions, or grading errors. 
An expert non-author may answer correctly, while a contributing author may struggle with questions that are outside their role. 
As such, an unsuccessful response does not by itself establish a lack of capacity to verify and should not be used categorically for determining a violation of submission policy. 
Instead, responsible development requires examining what a greCAPTCHA measures, the burdens it imposes on stakeholders, and how examinees' results could be interpreted and contested. 
As our initial investigation into its validity, we address two research questions:
\begin{description}[topsep=4pt]
    \item[\textbf{RQ1.}] How does researchers' performance on greCAPTCHA differ between their own and unfamiliar papers, and how do these differences vary across question families and domain familiarity?
    \item[\textbf{RQ2.}] How do researchers perceive the validity and acceptability of greCAPTCHAs, and what concerns do they identify about question quality, grading, deployment, and potential consequences?
\end{description}

\subsection{Evaluating The First Prototype greCAPTCHA}
\label{ssec:intro:grecaptcha}

We explore these research questions by running the first in-person greCAPTCHA with academics.

First, we design and implement a new prototype system which uses GenAI (GPT-5.6 Sol) to dynamically generate questions based on an author's manuscript and their stated contributions. In our prototype, authors read and respond to questions through a text interface, though an audio or video interface could be used in the future.
Informed by the revised Bloom's taxonomy of educational objectives~\cite{krathwohlRevisionBloomsTaxonomy2002}, we design four question families which target factual, conceptual, and procedural knowledge through tasks that ask participants to identify errors, explain choices, demonstrate background understanding, and assess limitations. 
The corresponding question families are: (1) planted-error; (2) unstated rationale; (3) background knowledge; (4) failure mode.

We evaluate greCAPTCHA with a mixed-methods study asking 31 researchers each to complete a one-hour in-person session. 
In a counterbalanced within-subjects design, participants answered questions about one of their own papers and an unfamiliar paper selected by the research team. 
Unfamiliar papers were selected to be either within or outside participants' research fields, allowing us to examine the role of domain familiarity.
Each assessment comprised eight questions, with two from each family, and a 15-minute response period. 
We then conducted semi-structured interviews to explore participants' experiences and their views on the validity, acceptability, and potential consequences of greCAPTCHAs.
We report ROC AUC to quantify how well greCAPTCHA scores separate participants' own papers from unfamiliar papers across all possible score thresholds, without committing to a particular deployment cutoff.

Our main \emph{quantitative} findings are:
\begin{itemize}
    \item \textbf{High authorship separation:} Our prototype distinguishes performance on participants' own and unfamiliar papers with a high AUC of $0.90$.
    \item \textbf{No statistically detectable field effect:} We find no statistically significant difference between scores on in-field and out-of-field unfamiliar papers overall.
\end{itemize}
Our main \emph{qualitative} findings are:
\begin{itemize}
    \item \textbf{Effective questions target less consequential knowledge:} Participants highlighted that deep questions about minor details are the best at providing evidence about capacity to verify in contrast to overall motivation or significance.
    \item \textbf{Assessment can promote learning and preparation:} Questions prompted reflection, a desire to learn, and even new insights for participants, some adding that it would be a valuable preparation aid for presentations. 
    In some cases, disagreement with the system led participants to doubt their expertise---including the judgment needed to challenge the assessment.
    \item \textbf{Grading requires careful calibration:} Participants highlighted that grading and feedback requires careful design beyond the current prototype. The expected level of detail should be clear and the grader should be flexible if a participant gives an alternative but equally correct answer.
    \item \textbf{Participants support deployment but do not agree on role:} Participants largely supported deployment of greCAPTCHAs and offered insights on potential decision mechanisms, such as automated desk-rejection, score-based badges, and tiered screening. In addition, human oversight and accessible recourse mechanisms were broadly requested features.
\end{itemize}
These findings provide a promising initial empirical basis for further investigating greCAPTCHA assessments as a source of evidence about authors' understanding and accountability.  
The code to run our prototype implementation of greCAPTCHA as well as the anonymized responses of the $31$ participants are available at \url{https://github.com/justinpayan/greCAPTCHA}.


\section{Background and Related Work}
\label{sec:background}

We ground our research by first establishing what constitutes authorship in order to understand how changing practices of GenAI-enabled scientific writing impact the accountability of authors.   
We then review how academic institutions and venues are responding to these changes and how those responses can go awry.
This information contextualizes our subsequent discussion of verification-relevant understanding and the automation of its measurement.

\subsection{Authorship and Submission Policies}

Scientific authorship has many definitions (see, for example~\cite{ACMPolicyAuthorship,ICMJERecommendationsDefining}). While the exact details may differ among disciplines and venues, most criteria agree that authors are expected to provide accountable human judgment, not merely written content.
Accountability is fundamental to greCAPTCHAs, not least because there are legitimate use cases for GenAI in scientific research.
GenAI is often used for language assistance for non-native English speakers~\cite{gartenbergMoreBetterArtificial2026,liangGPTDetectorsAre2023,liuAIAssistedWritingGrowing2025,huNegotiatingIdentityAge2025}.
In addition, applications such as revisions and proofreading~\cite{ratkovicHarnessingGPTEnhanced2025,wangHumanAICollaborationScience2026,liangCanLargeLanguage2023}, qualitative coding~\cite{gao2024collabcoder,parfenovaTextAnnotationInductive2025,qiaoGenerativeAIThematic2025}, literature exploration~\cite{asaiSynthesizingScientificLiterature2026,radenskyHumanLLMCompoundSystem2026,skarlinskiLanguageAgentsAchieve2024,nursalUnderstandingLLMsSummarization2026}, and accessibility~\cite{august2023paper,chheda-kotharyQueryingMultimodalScientific2026} can be key reasons why GenAI should be allowed in scientific research.

\begin{table}[t]
    \centering
    \caption{Four overlapping strands of policies governing GenAI use in research submission and review. Summaries describe common rules, but requirements may vary across venues, publishers, and professional bodies.}
    \label{tab:policies}

    \begin{tabular}{@{}P{0.125\linewidth}P{0.5\linewidth}P{0.32\linewidth}@{}}
    \toprule
    \textbf{Category} & \textbf{Policy summary} & \textbf{Example venues} \\
    \midrule
    
    \textit{Authorship \&} \newline \textit{accountability}
    & AI systems cannot be authors. Human authors remain responsible for all content. Scientific reasoning, interpretation, and judgment may not be delegated to GenAI.
    & Science~\cite{thorp2023chatgpt}; ICLR~\cite{iclr2025reviewer}; ICML~\cite{icml2023llm}; COPE~\cite{cope2023authorship}; ICMJE~\cite{icmje2026authors} \\
    \addlinespace
    
    \textit{Disclosure \&} \newline \textit{verification}
    & Substantive GenAI use must be identified, though routine language editing may be exempt. Authors must verify generated claims, references, code, and analyses. Nondisclosure or unverified generation can lead to rejection or post-publication action.
    & ACL~\cite{acl2023aiwriting}; NeurIPS~\cite{neurips2025llm}; CVPR~\cite{cvpr2025authors}; IEEE~\cite{ieee2024ai}; BMJ~\cite{bmjAIuse}; JAMA~\cite{flanagin2026guidance}; USENIX~\cite{usenix2026genai} \\
    \addlinespace
    
    \textit{Research evidence}
    & GenAI may not fabricate or materially alter primary research images, data, or results. GenAI use in methods and for some explanatory graphics may be permitted when disclosed, validated, accurately captioned, and reproducible.
    & Nature~\cite{natureAIpolicy}; Elsevier~\cite{elsevierAIpolicy}; Sage~\cite{sageAIpolicy}; Taylor \& Francis~\cite{taylorFrancisAIpolicy}; ACS~\cite{acs2024ai}; AIP~\cite{aipAIpolicy} \\
    \addlinespace
    
    \textit{Review} \newline \textit{integrity}
    & Reviewers may not upload confidential submissions to unapproved GenAI systems or delegate evaluation to GenAI. Limited language or background assistance may be allowed. Hidden prompts to manipulate AI-assisted review are misconduct.
    & ACL Rolling Review~\cite{aclarr2025reviewer}; ICML~\cite{icml2026llm}; ICCV~\cite{iccv2025reviewer}; CVPR~\cite{cvpr2026reviewer}; PLOS~\cite{plosAIpolicy}; IOP~\cite{iopAIauthors,iopAIreviewers}  \\
    \bottomrule
    \end{tabular}
\end{table}

While the use of GenAI for scientific research is not problematic by itself, its use does introduce a gap between authors and manuscripts which makes the use of text as evidence for authorship no longer admissible.
This realization is partly what leads numerous venues to re-evaluate their submission policies.
We organize these responses into four overlapping strands in~\cref{tab:policies}: authorship and accountability, disclosure and verification, research evidence, and review integrity. 
For greCAPTCHAs, the central question is how these responsibilities can be assessed: what evidence supports judgments about compliance, and what can that evidence establish? 
Examining these strands helps us position manuscript-specific assessments within the broader effort to maintain research integrity.

The first strand, \textbf{authorship and accountability}, assigns responsibility for submitted work to human authors and establishes boundaries on the delegation of intellectual work to GenAI~\cite{thorp2023chatgpt,iclr2025reviewer,icml2023llm,cope2023authorship,icmje2026authors}. 
Authorship declarations, in which authors check a box or sign their name to indicate they accept the responsibilities of authorship, provide a record of an author accepting these responsibilities but not evidence of the understanding needed to fulfill them. 
Investigating suspected violations requires editors to separate legitimate assistance from inappropriate delegation, often without direct access to the research process. 
greCAPTCHAs could complement declarations by eliciting evidence of authors' understanding of their stated contributions. 
Such evidence would inform the assessment of their capacity to take responsibility, without establishing who performed the original work.

The second strand, \textbf{disclosure and verification}, requires authors to report relevant GenAI use and check generated content, including claims, references, code, and analyses~\cite{acl2023aiwriting,neurips2025llm,cvpr2025authors,ieee2024ai,bmjAIuse,flanagin2026guidance,usenix2026genai}. 
Disclosures are used here to report tool use, and examining a manuscript can also reveal errors that warrant further investigation.
However, neither establishes whether an author possesses the knowledge needed to evaluate the content.
This distinction directly motivates our construct of capacity to verify because manuscript-specific questions may elicit evidence of the comprehension and reasoning needed for verification. 

The third strand, \textbf{research evidence}, governs GenAI use in producing or modifying research materials and seeks to preserve the authenticity and traceability of scientific findings~\cite{natureAIpolicy,elsevierAIpolicy,sageAIpolicy,taylorFrancisAIpolicy,acs2024ai,aipAIpolicy}. 
Raw data, metadata, code, and provenance records can support investigations, though maintaining and inspecting them imposes documentation and inspection costs on authors and evaluators. 
This strand clarifies an important boundary of greCAPTCHA. 
Questions about methodological choices and limitations assess understanding, but cannot authenticate data or establish reproducibility. 
A knowledgeable author can still fabricate evidence. 
Assessments of understanding would therefore complement, rather than replace scrutiny of the underlying research materials.

The fourth strand, \textbf{review integrity}, protects confidentiality and human judgment in evaluation and prohibits manipulation of AI-assisted reviewing~\cite{aclarr2025reviewer,icml2026llm,iccv2025reviewer,cvpr2026reviewer,plosAIpolicy,iopAIauthors,iopAIreviewers}. 
Although our study examines authors rather than reviewers, these policies also bear on the design of greCAPTCHAs: an assessment that uses GenAI to generate questions and evaluate responses must itself support trustworthy evaluation. 
In particular, generated questions and grades require scrutiny, manuscript confidentiality requires protection, and affected authors need meaningful ways to contest errors. 
Automating an assessment does not remove institutional responsibility for its quality or consequences.

Together, these strands distinguish requirements for responsible authorship from the evidence available to assess them. 
greCAPTCHAs address one part of this problem by eliciting manuscript-specific understanding relevant to authors' capacity to verify their contributions. 
Their usefulness depends on whether responses provide valid evidence of that capacity, how assessment errors and opportunities for gaming are handled, and who bears the resulting burden. 

\subsection{Designing Assessments of the Capacity to Verify}
\label{ssec:background:validity}

The previous policies establish a need for evidence about authors' understanding, but leave open how to elicit and interpret that evidence.
We approach the design of greCAPTCHAs as a form of \emph{assessment}, motivated by established practices such as oral examinations, thesis defenses, and interactive questioning. 
These practices ask people to explain and critically examine work for which they claim responsibility. 
Accordingly, greCAPTCHA seeks evidence of manuscript-specific understanding and judgment through participants' responses to questions about their work.

Following~\textcite{dawsonValidityMattersMore2024}, the validity of an assessment concerns whether evidence supports a particular \emph{interpretation and use} of performance. 
Our intended interpretation concerns \emph{capacity to verify}: the knowledge and reasoning needed to critically assess the claims, methods, and evidence underlying one's contributions to a manuscript.

Our construct is based on the concept of \emph{evaluative judgment}~\cite{taiDevelopingEvaluativeJudgement2018}.
This framework states that two main components must be present in order for someone to make decisions about the quality of their work: (1) understanding what constitutes quality; and (2) applying this understanding through an appraisal of work.
Applied to manuscript assessment, this perspective motivates questions that elicit both comprehension of the work and justification of its underlying choices. 
We therefore ask participants to demonstrate relevant knowledge and use it to explain decisions, identify problems, and assess limitations.

\begin{table*}[t]
    \centering
    \caption{The four question families used in the greCAPTCHA evaluation, including response formats, descriptions, and example questions.}
    \label{tab:question-categories}
    \small
\setlength{\tabcolsep}{5pt}
    \begin{tabularx}{\textwidth}{@{}
        P{0.12\textwidth}
        P{0.1\textwidth}
        P{0.35\textwidth}
        >{\RaggedRight\arraybackslash\hspace{0pt}}X
    @{}}
    \toprule
    \textbf{Question} \newline \textbf{family} &
    \textbf{Response format} &
    \textbf{Description} &
    \textbf{Example} \\
    \midrule

    \textit{Planted-error detection} &
    Multiple choice &
    Identify which version of a specific, verifiable claim accurately reflects the manuscript. &
    \emph{``In validating the three-item ethical-concern scales, which construct had a Cronbach's $\alpha$ of 0.82? A) Autonomy; B) Privacy; C) Fairness; D) Transparency.''} \\
    
    \addlinespace
    \textit{Unstated} \newline \textit{rationale} &
    Free-response &
    Explain why a methodological or design choice is appropriate, where the reason for that choice is not explicitly stated in the paper. &
    \emph{``Table 1 places both sensing-not-in-use scenarios before all four sensing-in-use scenarios, while randomising only within those blocks. What supports this fixed ordering rather than randomising all six scenarios, and what interpretive cost does that choice create?''} \\
    
    \addlinespace
    \textit{Background knowledge} &
    Free-response &
    Define an important concept that the manuscript presupposes but does not define, and explain why that concept matters for understanding the reported work. &
    \emph{``In your own words, explain the Mann–Whitney rank-sum procedure, including what it tests and the observation-level assumption it ordinarily makes. Then explain why this procedure matters for the condition comparisons reported.''} \\
    
    \addlinespace
    \textit{Failure mode} &
    Free-response &
    Name a realistic, manuscript-specific condition under which the proposed method or central finding would degrade, and explain the mechanism producing that degradation. &
    \emph{``Name one realistic learning condition for which the reported preference for system-generated hints over teacher assistance might weaken or reverse, and explain why.''} \\
    
    \bottomrule
    \end{tabularx}
\end{table*}

Specifically, we use the revised Bloom's taxonomy to guide the knowledge and cognitive demands of the assessment~\cite{krathwohlRevisionBloomsTaxonomy2002}. 
For this prototype, we focus on factual, conceptual, and procedural knowledge. 
We adapt these dimensions to the manuscript context as follows:
\begin{itemize}[topsep=4pt]
    \item \textbf{Factual knowledge:} relevant terminology, specific details, and reported findings.
    \item \textbf{Conceptual knowledge:} relationships among the manuscript's claims, underlying concepts, assumptions, and supporting evidence.
    \item \textbf{Procedural knowledge:} how methods and analyses were performed, and the conditions under which particular procedures are appropriate.
\end{itemize}
These dimensions inform four question families: planted-error, unstated rationale, background knowledge, and failure mode (see~\cref{tab:question-categories}). 
The families provide different ways to elicit knowledge and reasoning, and individual questions may draw on more than one knowledge dimension.
Questions are generated using both the manuscript and the participant's stated contributions to improve their relevance to the work for which the participant claims responsibility.
Our first empirical analysis explores whether greCAPTCHA can distinguish author responses from non-author responses at the participant level.
This comparison provides initial evidence about assessment performance and construct validity, but its interpretation also depends on how questions are generated, how responses are graded, and how participants experience the assessment. 
We next consider the implications of automating these processes.

\subsection{Automating Authorship Assessment}
\label{ssec:background:automated-assessment}

The greCAPTCHA assessment approach outlined above requires questions tailored to individual manuscripts and authors' contributions.
Producing and grading these questions manually would demand subject-matter expertise and could add substantial labor to existing evaluation processes. 
Automated question generation offers a potential means of supporting this work~\cite{mitkovComputerAidedGeneration2003,heilmanGoodQuestionStatistical2010,duLearningAskNeural2017,kurdiSystematicReviewAutomatic2020}. 
For greCAPTCHAs, however, the challenge extends beyond generating fluent questions: questions must be relevant to the author's contributions, elicit the intended knowledge and reasoning, and support defensible judgments about responses.

A closely related approach is the AI-mediated \textit{viva voce} assessment developed by~\textcite{EBRAHIMZADEH2026100609}. 
Their system, VivaBuddy, uses GenAI to generate comprehension questions and provide dialogic feedback about students' work. 
The authors introduce \emph{coauthorship integrity}, which concerns students' understanding of work produced in collaboration with GenAI. 
This approach shares our motivation to elicit understanding rather than infer it from the presence or absence of AI-generated text. 
Our proposed assessment extends their approach to all forms of research manuscripts where specialized subject matter, shared contributions, and professional responsibility shape what authors can be expected to explain. 

Using GenAI to generate questions and grade responses introduces additional sources of uncertainty. 
Generated questions may contain mistaken premises, introduce material outside an author's contribution, or admit reasonable answers that an automated grader fails to recognize. 
Consequently, differences in scores may reflect properties of the assessment as well as differences in understanding. 
We treat automated scores as evidence whose interpretation depends on question relevance, grading criteria, and assessment conditions. 

The potential consequences of automated assessment also require attention to fairness and legitimacy~\cite{dawsonValidityMattersMore2024}. 
Language proficiency, career stage, disability, anxiety, and unequal testing conditions may all influence what participants can demonstrate. 
Proctoring introduces further questions concerning privacy, security, accessibility, and cost~\cite{hutchinsonReExaminingExaminers2026}. 
Any consequential deployment would therefore require consideration of accommodations, scrutiny of automated judgments, and meaningful opportunities to contest questions, grades, and resulting decisions. 
Work on algorithmic decision-making and contestability highlights the importance of these procedures to how people experience and evaluate automated judgments~\cite{binnsItsReducingHuman2018,karusalaUnderstandingContestability2024}.

We therefore approach greCAPTCHAs as a sociotechnical assessment process whose usefulness depends on both the evidence elicited and the conditions under which it is interpreted and used. 
This perspective motivates RQ2: examining researchers' experiences and perceptions can reveal concerns about relevance, grading, burden, and legitimacy that performance comparisons alone could not identify. 
Our mixed-methods evaluation of greCAPTCHA brings these perspectives together to investigate the promise and limitations of an initial implementation.


\section{Assessing Understanding with greCAPTCHA}
\label{sec:sys}

We describe the general procedure of administering a greCAPTCHA, then discuss the specific prototype implementation used in our study.

\subsection{Design}
\label{ssec:sys:design}

The greCAPTCHA process translates the assessment approach described in~\cref{ssec:background:automated-assessment} into four stages: \textbf{submission}, \textbf{generation}, \textbf{examination}, and \textbf{reporting} (\cref{fig:greCAPTCHA-teaser}). 
We distinguish two roles: the \emph{examinee}, whose understanding is assessed, and the \emph{administrator}, who configures and administers the assessment. 

\paragraph{Submission.}
The examinee submits their manuscript and a statement describing their specific contributions. 
The manuscript provides the assessment content, while the contribution statement identifies the parts of the work for which the examinee claims responsibility. 
Both inputs guide question generation so that the assessment targets knowledge relevant to the examinee's role.
This process may vary across publication, coursework, and recruitment settings.

\paragraph{Generation.}
The system parses the manuscript and generates questions and corresponding grading rubrics using the contribution statement and the administrator's assessment configuration. 
This configuration specifies the question families, question counts, models, and prompts used for generation and grading. 
The administrator also establishes examination conditions, including time limits, response requirements, question ordering, and permitted resources. 
These choices should reflect the submitted work and the purpose of the assessment: the four question families evaluated in this paper are just one configuration of the broader greCAPTCHA approach.

\paragraph{Examination.}
The examinee answers the generated questions under the specified proctored conditions. 
The system presents the questions and records the responses for grading. 
In deployment, dedicated test centers are a possible venue for the assessment, while certified take-home exams---similar to those offered by Duolingo~\cite{DuolingoEnglishTest} and TOEFL~\cite{TOEFLIBTTesting}---could provide a lower-burden setup. Assessments might also be implemented at professional events, such as conferences, hackathons, or career fairs.

\paragraph{Reporting.}
On submission, the responses of the examinee are graded against the corresponding rubrics to produce scores. 
This process generates an assessment report that contains the participant's overall and per-question scores, their responses, how those responses matched the rubric, any partial-credit allocation, and feedback.
The report is bundled with the manuscript and the examinee's contribution statement and then shared with the requesting institution according to the exam arrangements.

\subsection{Implementation}
\label{ssec:methods:system}

We implement greCAPTCHA as a web application with two interfaces: an \textbf{administrator dashboard} and an \textbf{exam screen} for examinees.
The dashboard supports assessment preparation and management, while the exam screen presents the generated questions and collects responses. 
This separation allows administrators to configure assessments without requiring examinees to interact with the generation settings.

\begin{figure}[t]
    \centering
    \includegraphics[width=0.49\linewidth]{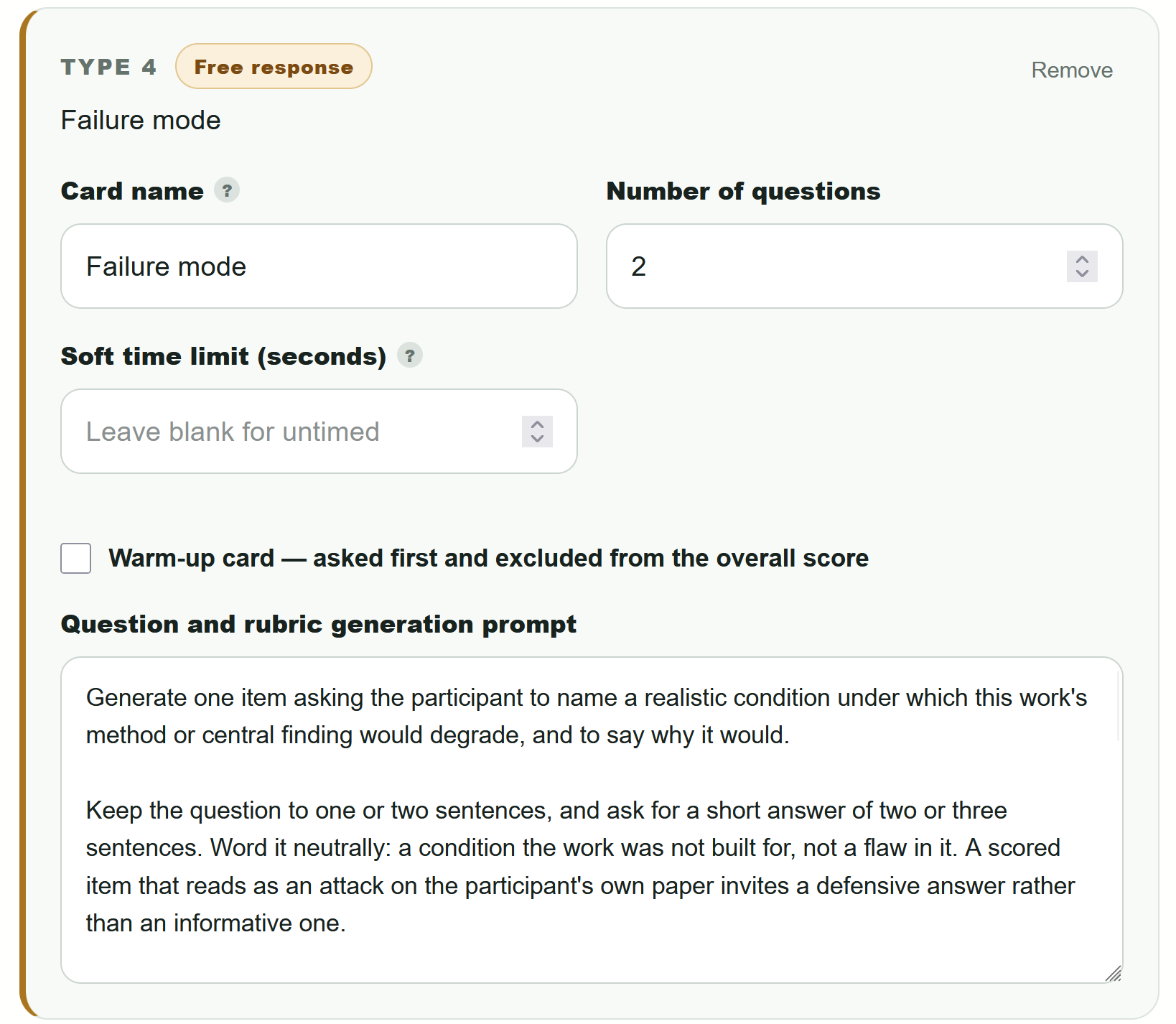}
    \hfill
    \includegraphics[width=0.49\linewidth]{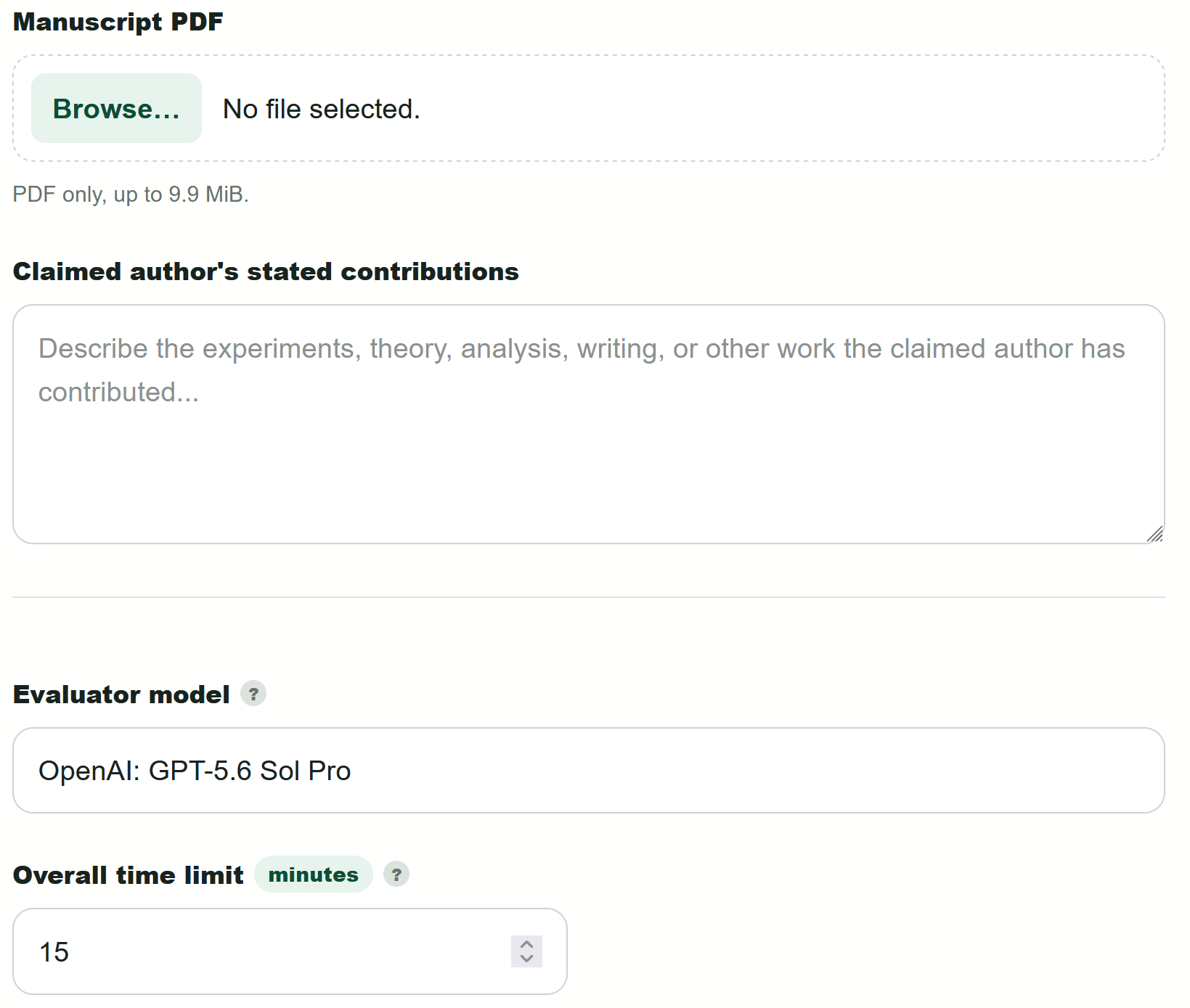}
    \caption{Screenshots from our greCAPTCHA implementation's administrator dashboard. The left image shows options for configuration of a question family, including its generation prompt and question settings. The right image shows manuscript upload, contribution statement, and assessment-level settings.}
    
    \label{fig:system-screenshots}
\end{figure}

\paragraph{Assessment preparation and management.}
An authenticated administrator uploads the examinee's manuscript and enters their contribution statement through the dashboard (\cref{fig:system-screenshots}). 
The administrator selects model settings and randomization options and configures the question families used to generate the assessment.
The dashboard supports a configurable number of families, each with its own generation prompt and question settings, such as the number of questions and, for multiple-choice items, the number of distractors. 
It also provides controls for timing and warm-up questions. 
The system supports multiple attempts on a question set, organization of question sets into sequential experiments, and question-set overviews.

\paragraph{Question families used in the study.}
For our evaluation, we configure four question families informed by the assessment rationale in~\cref{ssec:background:validity}: planted-error detection, unstated rationale, background knowledge, and failure mode. 
Their response formats and examples are presented in~\cref{tab:question-categories}.
The complete generation prompts appear in Appendix~\ref{apx:question-prompts}.

Unstated-rationale questions ask examinees to explain choices underlying the manuscript. 
Background-knowledge questions probe concepts needed to understand the work, while failure-mode questions ask examinees to reason about methodological limitations and circumstances in which an approach may fail. 
These three families use free-response formats. 
Planted-error detection uses a multiple-choice format, providing a structured task in which examinees identify the correct option among multiple incorrect options related to the manuscript.

The families are intended to elicit complementary evidence of understanding rather than isolate mutually exclusive types of knowledge.
Explaining a design choice, for example, may require knowledge of the manuscript, its domain, and the research process.
This overlap follows from the reasoning required by the questions, rather than from the free-response format alone. 
Planted-error detection provides a comparison with the free-response families, but performance on this task may mostly reflect attention.


\section{Evaluation of greCAPTCHA}
\label{sec:methods}

\subsection{Interview Protocol}

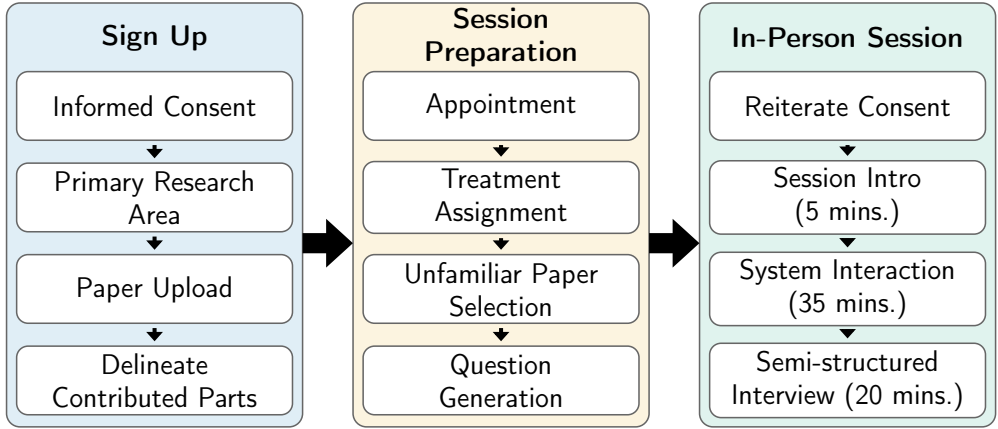
\begin{figure}
  \centering
  \resizebox{0.8\textwidth}{!}{

\begin{tikzpicture}[
  x=1cm,
  y=1cm,
  font=\normalfont\sffamily,
  workflow panel/.style={
    draw=black!60,
    line width=0.65pt,
    rounded corners=0.2cm
  },
  panel one/.style={workflow panel,fill=wfPanelI},
  panel two/.style={workflow panel,fill=wfPanelII},
  panel three/.style={workflow panel,fill=wfPanelIII},
  workflow box/.style={
    draw=black!60,
    fill=white,
    line width=0.65pt,
    rounded corners=0.17cm,
    minimum width=4.10cm,
    minimum height=\wfBoxH cm,
    align=center,
    inner xsep=2mm,
    inner ysep=1.0mm,
    font=\normalfont\sffamily\large          
  },
  workflow title/.style={
    align=center,
    font=\normalfont\sffamily\large    
  },
  workflow down arrow/.style={
    -{Triangle[width=2.4mm,length=1.4mm]},
    line width=0.6mm,
    shorten <=0.6mm,
    shorten >=0.6mm,
    draw=black
  },
  workflow across arrow/.style={
    -{Triangle[width=7mm,length=4.0mm]},
    line width=3.0mm,
    draw=black
  }
]

\definecolor{wfPanelI}{RGB}{224,238,246}   
\definecolor{wfPanelII}{RGB}{252,244,224}  
\definecolor{wfPanelIII}{RGB}{224,243,238} 

\pgfmathsetmacro{\wfBoxH}{1.02}   
\pgfmathsetmacro{\wfGap}{0.34}    
\pgfmathsetmacro{\wfTop}{1.02}    
\pgfmathsetmacro{\wfPad}{0.18}    
\pgfmathsetmacro{\wfRowA}{-\wfTop-0.5*\wfBoxH}
\pgfmathsetmacro{\wfRowB}{\wfRowA-\wfBoxH-\wfGap}
\pgfmathsetmacro{\wfRowC}{\wfRowB-\wfBoxH-\wfGap}
\pgfmathsetmacro{\wfRowD}{\wfRowC-\wfBoxH-\wfGap}
\pgfmathsetmacro{\wfBot}{\wfRowD-0.5*\wfBoxH-\wfPad}
\pgfmathsetmacro{\wfMid}{0.5*(\wfRowB+\wfRowC)}
\pgfmathsetmacro{\wfTitleY}{-0.5*\wfTop}

\draw[panel one]   (0,0)     rectangle (4.40,\wfBot);
\draw[panel two]   (5.15,0)  rectangle (9.55,\wfBot);
\draw[panel three] (10.30,0) rectangle (14.70,\wfBot);
\node[workflow title] at (2.20,\wfTitleY)  {\textbf{Sign Up}};
\node[workflow title] at (7.35,\wfTitleY)  {\textbf{Session}\\\textbf{Preparation}};
\node[workflow title] at (12.50,\wfTitleY) {\textbf{In-Person Session}};
\node[workflow box] (a1) at (2.20,\wfRowA) {Informed Consent};
\node[workflow box] (a2) at (2.20,\wfRowB) {Primary Research\\Area};
\node[workflow box] (a3) at (2.20,\wfRowC) {Paper Upload};
\node[workflow box] (a4) at (2.20,\wfRowD) {Delineate\\Contributed Parts};
\node[workflow box] (b1) at (7.35,\wfRowA) {Appointment};
\node[workflow box] (b2) at (7.35,\wfRowB) {Treatment\\Assignment};
\node[workflow box] (b3) at (7.35,\wfRowC) {Unfamiliar Paper\\Selection};
\node[workflow box] (b4) at (7.35,\wfRowD) {Question\\Generation};
\node[workflow box] (c1) at (12.50,\wfRowA) {Reiterate Consent};
\node[workflow box] (c2) at (12.50,\wfRowB) {Session Intro\\(5 mins.)};
\node[workflow box] (c3) at (12.50,\wfRowC) {System Interaction\\(35 mins.)};
\node[workflow box] (c4) at (12.50,\wfRowD) {Semi-structured\\Interview (20 mins.)};
\draw[workflow down arrow] (a1.south) -- (a2.north);
\draw[workflow down arrow] (a2.south) -- (a3.north);
\draw[workflow down arrow] (a3.south) -- (a4.north);
\draw[workflow down arrow] (b1.south) -- (b2.north);
\draw[workflow down arrow] (b2.south) -- (b3.north);
\draw[workflow down arrow] (b3.south) -- (b4.north);
\draw[workflow down arrow] (c1.south) -- (c2.north);
\draw[workflow down arrow] (c2.south) -- (c3.north);
\draw[workflow down arrow] (c3.south) -- (c4.north);
\draw[workflow across arrow] (4.40,\wfMid)  -- (5.15,\wfMid);
\draw[workflow across arrow] (9.55,\wfMid) -- (10.30,\wfMid);
\end{tikzpicture}}
  \caption{Our study workflow consists of three stages. The first stage requires participants to \textbf{Sign Up} using an online form, upload one of their own papers, and briefly explain which parts of the paper they contributed to. The second stage is for \textbf{Session Preparation} where we arrange appointments with each participant and assign them to experimental conditions. For the third stage we deliver the \textbf{In-Person Session} which involves a 5-minute introduction, a 35-minute interaction with greCAPTCHA, and a 20-minute semi-structured interview.}
  \label{fig:study-workflow}
\end{figure}

Given the above setup for greCAPTCHA, we recruit and conduct a mixed-methods evaluation with in-person participants.  
A timeline of a participant's experience from the start to end of the study is shown in~\cref{fig:study-workflow} and it has three stages: (1) Sign Up; (2) Session Preparation; and (3) In-Person Session.
The study was piloted with four members of the authors' research group.

In the first stage, participants first receive information about the study and provide informed consent.
Having consented to taking our study, participants then enter their primary research area, upload their \textit{own paper} on which they contributed substantially, and describe their contributions to the uploaded paper.

In the second stage, we schedule individual one-hour-long in-person sessions with each study participant via email. 
We then select an \textit{unfamiliar paper} to assign to them.
We standardize this selection process by prompting GPT-5.6 Sol with a prompt template (see Appendix~\ref{apx:paper-selection-prompt}) to suggest five potential papers from the preprint server arXiv.
As preprints can be low quality, we choose the paper which we judge highest quality based on an initial reading of the papers and the recognizability of their authors' institutions.
Participants are then assigned to one of four treatment conditions based on two independent variables (IV)---one within-subjects and one between-subjects:
\begin{itemize}[topsep=4pt]
    \item \textbf{Paper Authorship} [within subjects]: this binary IV has two conditions: (1) own paper and (2) unfamiliar paper.
    The own paper condition tests participants on questions based on their submitted manuscript and contribution statement. The unfamiliar paper condition tests authors on the unfamiliar paper that was assigned to them. For this condition, the question generation system is told that the participant ``did everything for the paper'' as the contribution statement. We counterbalance the order of the two conditions.
    \item \textbf{Unfamiliar Paper Field} [between subjects]: this binary IV determines whether a participant is assigned an unfamiliar paper that is closely related to their field (\textit{i.e.,} in-field) or that is distant from their field (\textit{i.e.,} out-of-field).
    Field familiarity in this case means that the participant's training transfers to the unfamiliar paper. They are familiar with the notation, methods, and background knowledge, and they could read the paper without learning a new apparatus.
\end{itemize}
This results in a $2\times2$ design, where each treatment condition is a unique combination of paper authorship order and unfamiliar paper field.
One day before the session, the selected unfamiliar papers are emailed to the participants.
Prior to the interview, we generate the question sets for both own and unfamiliar paper.
We generate eight questions, two from each of the four categories, and randomize the question order to avoid sequence effects.

In the third stage, during the study session, each participant is introduced to the system in $5$ minutes.
They are then instructed to interact with the greCAPTCHA system with a $15$-minute hard time limit per paper, for a maximum total of $35$ minutes, including $5$ minutes of overhead.
Participants are not told the exact purpose of the system beforehand.
These examination conditions are chosen to facilitate in-person interviews on an academic budget.
In realistic deployments a greCAPTCHA should involve more questions with a longer time limit to reduce variance.
It must also be calibrated to the particular context of the examinees, taking into account the fairness, accessibility, and privacy of the participants.
We record the following dependent variables:
\begin{itemize}[topsep=4pt]
    \item \textbf{Scores:} The score assigned by the automated grader to each of the participant's answers on both own and unfamiliar paper. The average score over all questions in a test is also a measure of interest.
    \item \textbf{Skipped:} Which questions were skipped by the participant.
\end{itemize}

Following the interaction, we conduct a $20$-minute-long semi-structured interview with the participant, covering their overall experiences with the system, their opinions about questions and evaluation quality, and their recommendations about whether and how to deploy the system in the broader academic system, including peer review and admissions.
The interview topics are summarized in \cref{tab:interview-themes}, and our full interview script is available at \url{https://github.com/justinpayan/greCAPTCHA}. 
Participants were then compensated $\$30$ via bank transfer (using Venmo/Zelle) for their time.
This study was approved by the Institutional Review Board of Carnegie Mellon University under study number \texttt{2026\_333}.

\begin{table}[t]
    \centering
    \caption{Post-interaction interview themes and representative follow-up probes.}
    \label{tab:interview-themes}
    
  \small
  \renewcommand{\arraystretch}{1.08}
  \setlength{\tabcolsep}{5pt}

  \begin{tabularx}{\linewidth}{
    @{}
    P{0.25\linewidth}
    >{\RaggedRight\arraybackslash}X
    @{}
  }
    \toprule
    \textbf{Theme} & \textbf{Representative follow-up probes} \\
    \midrule

    \textit{Overall reaction}
      & What stood out most or surprised you?
        \par\smallskip
        At what points, if any, were you uncertain about what the system expected? \\
    \addlinespace[0.65em]

    \textit{Grading and feedback}
      & Were any correct or reasonable answers marked incorrect, or weak answers accepted?
        \par\smallskip
        Did the explanations make the grading and supporting evidence understandable? \\
    \addlinespace[0.65em]

    \textit{Question quality}
      & Which questions provided the strongest or weakest evidence of understanding, and why?
        \par\smallskip
        Could someone answer correctly without understanding the paper, or understand it yet answer incorrectly? \\
    \addlinespace[0.65em]

    \textit{Paper-specific assessment}
      & For your paper, did the questions reflect the portions you contributed to or verified?
        \par\smallskip
        For the unfamiliar paper, did the system distinguish field knowledge from familiarity with that specific paper? \\
    \addlinespace[0.65em]

    \textit{Perceived purpose and construct validity}
      & What did the system appear to measure: understanding, verification, personal contribution, or something else?
        \par\smallskip
        Did it unintentionally measure memory, test-taking ability, English fluency, or speed? \\
    \addlinespace[0.65em]

    \textit{Circumvention and strategic behavior}
      & How might authors prepare for or work around the system without verifying their papers?
        \par\smallskip
        Which parts would be hardest to game, and would countermeasures create privacy, fairness, or usability concerns? \\
    \addlinespace[0.65em]

    \textit{Trust and deployment}
      & How would using such a system affect your trust in a conference or journal?
        \par\smallskip
        What role, if any, should it have when it is uncertain or appears to make a mistake? \\
    \addlinespace[0.65em]

    \textit{Recommendation}
      & Would you recommend that a colleague's conference adopt the system?
        \par\smallskip
        What single change would make it more acceptable or effective? \\

    \bottomrule
  \end{tabularx}
\end{table}

\subsection{Participants}
\label{ssec:methods:participants}

\begin{figure}
    \centering
    \input{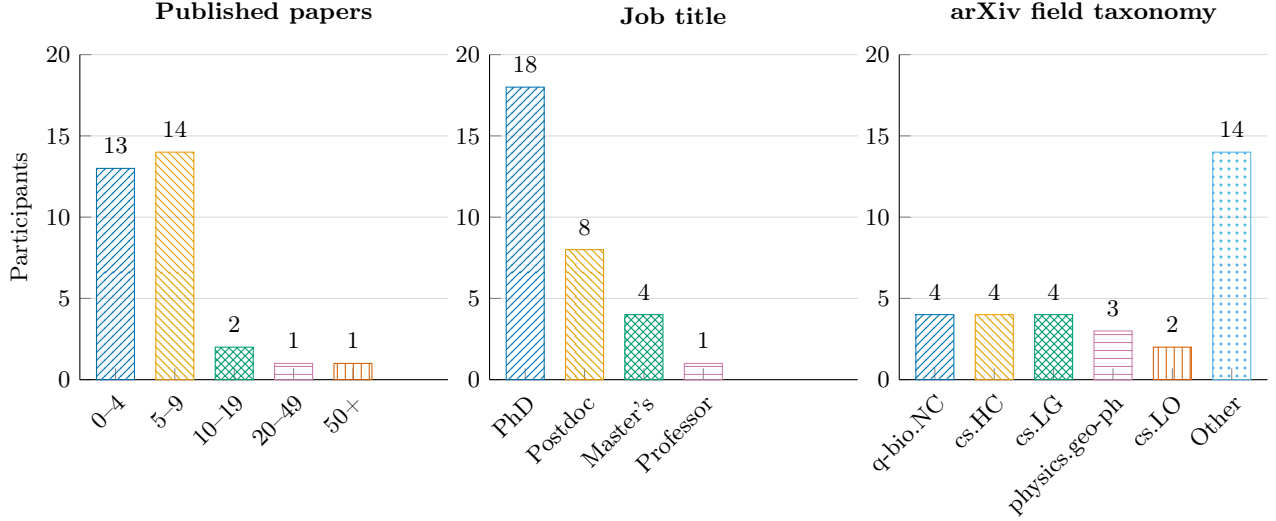}
    \vspace{-2em}
    \caption{Participant characteristics ($n=31$).
    arXiv fields: q-bio.NC = Neurons and Cognition; cs.HC = Human-Computer Interaction; cs.LG = Machine Learning; physics.geo-ph = Geophysics; cs.LO = Logic in Computer Science.}
    \label{fig:participant-characteristics}
\end{figure}

A total of $31$ participants completed the study.
Subjects were recruited via email, social media, message boards, in-person solicitation, and through word-of-mouth from the academic and research institutions in Pittsburgh, PA, USA.
A distribution of their relevant background information is shown in~\cref{fig:participant-characteristics} with the full underlying information displayed in Appendix~\ref{apx:participants}.
Participants primarily came from cognitive science and computing disciplines, with at least one each from philosophy, psychology, hydrology, and environmental engineering.
Most participants were PhD students, followed by postdoctoral researchers (or equivalent industry positions), and the majority of our participants had fewer than $10$ public manuscripts.
Participants were assigned equally at random to one of our four treatment conditions, giving eight participants per condition, except for the own-paper-first out-of-field condition, which received $7$ participants.

\subsection{Statistical Methods}
\label{ssec:methods:quant}

We address RQ1 by computing the Receiver Operating Characteristic (ROC) curve and calculating the Area Under this Curve (AUC) from the overall scores assigned to participants. 
The ROC is especially helpful to understand how well our greCAPTCHA separates true positives (\textit{i.e.,} non-authors classified as non-authors) from false positives (\textit{i.e.,} authors misclassified as non-authors).
We also view the distribution of scores assigned within each question family, and construct an ROC curve and AUC for each question family.

To explore the effect of familiarity with the unfamiliar paper's field, our second analysis is restricted to responses only about unfamiliar papers.
Given these data, we examine whether performance differs between participants assigned to the in-field and out-of-field conditions.
We normalize scores to the range $[0, 1]$ by dividing by $100$. 
Because scores are bounded and clustered around $0$ and $1$ for the planted-error questions, we fit an ordered-beta mixed-effects regression model to our data~\cite{kubinec2023ordered}.
This model predicts authors' scores from fixed effects for the unfamiliar-paper field and question family, and includes random intercepts for each participant: \texttt{score $\sim$ field\_type + question\_family + (1 | participant\_id)}.
We estimate the contrast between in-field and out-of-field performance, testing the null hypothesis that this contrast is zero. Statistical modeling and figure-generation code were generated using GPT-5.6 Sol and verified by the first authors.

\subsection{Qualitative Methods}
\label{ssec:methods:qual}

The individual scores of participants measure their particular performance on our question sets under specific conditions.
We contextualize these scores by eliciting participants' views about their experience with greCAPTCHA.

To do this, we make audio recordings of each semi-structured interview.
We transcribe and diarize each recording using the WhisperX system~\cite{bain2022whisperx}.
We further preprocess the transcripts using deterministic rules and a language model constrained to removing only excessive usage of filler words, fixing punctuation, and to align the audio timestamps with speaker boundaries.
The preprocessed transcripts were then verified against the raw text to ensure that no words were invented or silently altered.
Speaker boundaries are manually verified against the recordings by one of the authors when it is ambiguous who is speaking.
All deep learning models were run on a local machine.

We perform a hybrid deductive–inductive codebook thematic analysis, organizing our codebook with the open-source software QualCoder~\cite{curtain_qualcoder_2025}.
First, two authors independently developed deductive codebooks based on the interview themes in~\cref{tab:interview-themes} and on skimming the transcripts; these were then consolidated into a single codebook.
In addition, we also perform inductive coding.
Once data collection finished, the same two authors independently coded the same four randomly selected interviews, then discussed the applications of inductive codes.
Finally, the authors merged a combined inductive codebook with their deductive codes.
This final codebook, along with descriptions of codes and annotation counts, is available in Appendices~\ref{apx:deductive_codebook} and \ref{apx:inductive_codebook}.

One author then coded all $31$ interview transcripts---recoding the previously selected four transcripts.
The first authors then collaboratively organized the coded passages into candidate themes and refined the themes through discussion.
We connect these themes back to our quantitative analyses by organizing participants' data into cases and analyzing their distribution along our identified themes.
This involves examining participants' accounts within each theme alongside their own-paper and unfamiliar-paper scores and their unfamiliar-paper field assignment.
This helps us identify explanations for observed performance patterns as well as analyze cases of disagreement.


\section{Quantitative Results}
\label{sec:quantitative}

Recall from~\cref{ssec:background:validity} that our quantitative analysis set out to explore two related questions.
First, we examine the trade-off between correctly identifying non-author responses and incorrectly classifying author responses as coming from non-authors.
Second, we look at how domain familiarity may affect the results of greCAPTCHA.

\subsection{Author Separation Accuracy}
\label{ssec:quantitative:classification_accuracy}

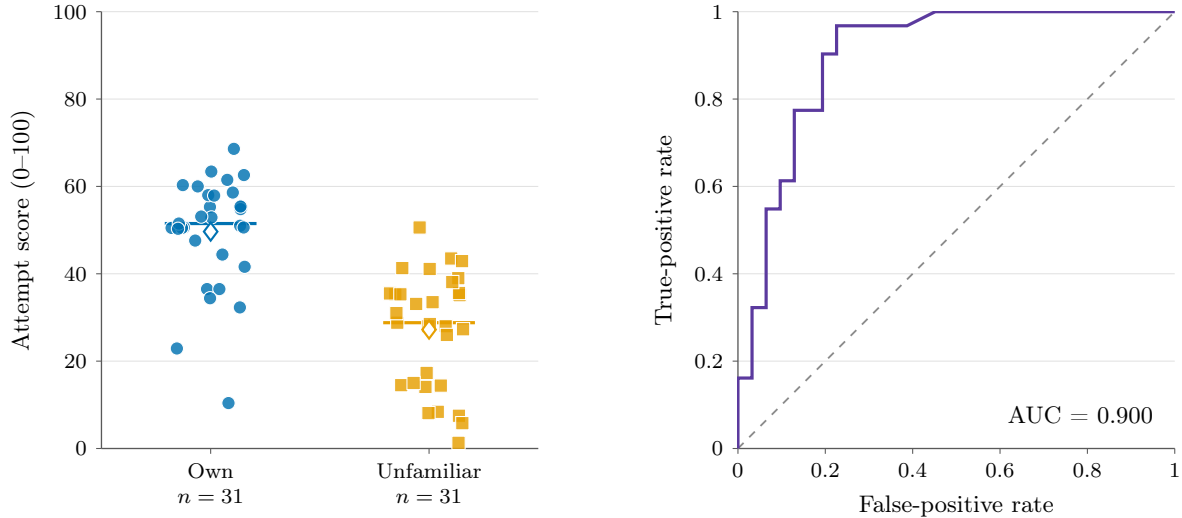
\begin{figure}
    \centering
\begingroup
\ifdefined\tikzexternaldisable\tikzexternaldisable\fi
\definecolor{ownColor}{HTML}{0072B2}
\definecolor{foreignColor}{HTML}{E69F00}
\definecolor{rocColor}{HTML}{5B3A9E}
\begin{tikzpicture}
\begin{groupplot}[
group style={group size=2 by 1,horizontal sep=0.16\linewidth},
scale only axis,
width=0.35\linewidth,height=0.35\linewidth,
axis lines*=left,axis line style={black!65,line width=0.5pt},
tick align=outside,tick style={black!65,line width=0.4pt},
major tick length=2pt,scaled ticks=false,ymajorgrids=true,
grid style={black!12,line width=0.35pt},
label style={font=\small},tick label style={font=\footnotesize},
]
\nextgroupplot[xmin=-0.5,xmax=1.5,ymin=0,ymax=100,xtick={0,1},xticklabels={{Own\\$n=31$},{Unfamiliar\\$n=31$}},xticklabel style={align=center},ytick={0,20,40,60,80,100},ylabel={Attempt score (0--100)},]
\addplot[only marks,mark=*,mark size=2.5pt,draw=white,fill=ownColor,fill opacity=0.82] coordinates {(0.1337599186,32.3) (0.1555939245,41.6) (-0.1546539216,22.9) (-0.1789549723,50.5) (0.1398379144,55.3) (-0.003337570861,55.3) (-0.01661197459,36.5) (0.1370026076,54.8) (-0.1226921527,50.6) (-0.1313463344,50.5) (0.03988443515,36.5) (0.002905215482,63.4) (0.1014061441,58.6) (0.1348124971,51) (0.1061980853,68.6) (-0.01110139679,58) (0.07603682441,61.5) (-0.145628171,51.5) (-0.127487364,60.3) (0.151034653,50.6) (0.08160111354,10.4) (0.01603691772,57.9) (0.05375019782,44.4) (0.1367208317,55.4) (0.003631968562,52.9) (-0.05917234476,60) (-0.002721842548,34.4) (0.152294548,62.6) (-0.1494982331,50.3) (-0.0707902145,47.6) (-0.04326253534,53.1)};
\addplot[forget plot,only marks,mark=diamond*,mark size=3.4pt,draw=ownColor,fill=white,line width=0.8pt] coordinates {(0,49.65483871)};
\draw[draw=ownColor,line width=1.2pt] (axis cs:-0.21,51.5) -- (axis cs:0.21,51.5);
\addplot[only marks,mark=square*,mark size=2.5pt,draw=white,fill=foreignColor,fill opacity=0.82] coordinates {(1.133759919,39) (1.155593925,27.3) (0.8453460784,35.3) (0.8210450277,35.5) (1.139837914,35.1) (0.9966624291,28) (0.9833880254,14.1) (1.137002608,35.6) (0.8773078473,41.3) (0.8686536656,35.3) (1.039884435,8.4) (1.002905215,41.1) (1.101406144,43.5) (1.134812497,1.3) (1.106198085,38.1) (0.9888986032,17.3) (1.076036824,28) (0.854371829,28.8) (0.872512636,14.5) (1.151034653,42.9) (1.081601114,26) (1.016036918,33.5) (1.053750198,14.4) (1.136720832,7.5) (1.003631969,28.5) (0.9408276552,33.1) (0.9972781575,8.1) (1.152294548,5.8) (0.8505017669,31) (0.9292097855,15) (0.9567374647,50.6)};
\addplot[forget plot,only marks,mark=diamond*,mark size=3.4pt,draw=foreignColor,fill=white,line width=0.8pt] coordinates {(1,27.22258065)};
\draw[draw=foreignColor,line width=1.2pt] (axis cs:0.79,28.8) -- (axis cs:1.21,28.8);
\nextgroupplot[xmin=0,xmax=1,ymin=0,ymax=1,xtick={0,0.2,0.4,0.6,0.8,1},ytick={0,0.2,0.4,0.6,0.8,1},xlabel={False-positive rate},ylabel={True-positive rate}]
\addplot[forget plot,black!45,dashed,line width=0.7pt] coordinates {(0,0) (1,1)};
\addplot[forget plot,rocColor,line width=1.2pt] coordinates {(0,0) (0,0.03225806452) (0,0.06451612903) (0,0.09677419355) (0,0.1290322581) (0,0.1612903226) (0.03225806452,0.1612903226) (0.03225806452,0.1935483871) (0.03225806452,0.2258064516) (0.03225806452,0.2580645161) (0.03225806452,0.2903225806) (0.03225806452,0.3225806452) (0.06451612903,0.3225806452) (0.06451612903,0.3548387097) (0.06451612903,0.3870967742) (0.06451612903,0.4516129032) (0.06451612903,0.4838709677) (0.06451612903,0.5161290323) (0.06451612903,0.5483870968) (0.09677419355,0.5483870968) (0.09677419355,0.5806451613) (0.09677419355,0.6129032258) (0.1290322581,0.6129032258) (0.1290322581,0.6451612903) (0.1290322581,0.7096774194) (0.1290322581,0.7419354839) (0.1290322581,0.7741935484) (0.1935483871,0.7741935484) (0.1935483871,0.8064516129) (0.1935483871,0.8387096774) (0.1935483871,0.8709677419) (0.1935483871,0.9032258065) (0.2258064516,0.9032258065) (0.2258064516,0.935483871) (0.2258064516,0.9677419355) (0.2580645161,0.9677419355) (0.2903225806,0.9677419355) (0.3225806452,0.9677419355) (0.3870967742,0.9677419355) (0.4516129032,1) (0.4838709677,1) (0.5161290323,1) (0.5483870968,1) (0.5806451613,1) (0.6129032258,1) (0.6774193548,1) (0.7096774194,1) (0.7419354839,1) (0.7741935484,1) (0.8064516129,1) (0.8387096774,1) (0.8709677419,1) (0.9032258065,1) (0.935483871,1) (0.9677419355,1) (1,1)};
\node[anchor=south east,font=\small,fill=white,inner sep=3pt] at (axis cs:0.97,0.04) {AUC = 0.900};
\end{groupplot}
\end{tikzpicture}
\Description{A two-panel figure. The left panel shows participant attempt scores from familiar own papers as blue circles in the left column and unfamiliar foreign papers as orange squares in the right column, on a shared score axis from 0 to 100. Small deterministic horizontal jitter reduces overplotting; points are not connected. Axis labels report n. White diamonds mark condition means and colored horizontal segments mark medians. Familiar scores have mean 49.7 and median 51.5; unfamiliar scores have mean 27.2 and median 28.8. The right panel is the empirical receiver operating characteristic curve obtained by using one minus the attempt score proportion to classify foreign versus own tests, with foreign as the positive class. The area under the curve is 0.900. The diagonal denotes chance-level discrimination. These are descriptive summaries and do not establish causation.}
\ifdefined\tikzexternalenable\tikzexternalenable\fi
\endgroup
    \caption{The left plot shows overall score distribution broken down by whether the paper tested is the participant's own paper or the unfamiliar paper ($n=31$ per category). Lines indicate medians and diamonds indicate means. On the right, we plot the ROC for prediction of ``unfamiliar'' label, using $(1-\text{score}/100)$ as the predictor.}
    \label{fig:participant_condition_score_scatterplots}
    
\end{figure}

\begin{figure}
    \centering
\begingroup
\ifdefined\tikzexternaldisable\tikzexternaldisable\fi
\definecolor{ownColor}{HTML}{0072B2}
\definecolor{foreignColor}{HTML}{E69F00}
\definecolor{rocColor}{HTML}{5B3A9E}
\begin{tikzpicture}
\begin{groupplot}[
group style={group size=2 by 1,horizontal sep=0.16\linewidth},
scale only axis,
width=0.35\linewidth,height=0.35\linewidth,
axis lines*=left,axis line style={black!65,line width=0.5pt},
tick align=outside,tick style={black!65,line width=0.4pt},
major tick length=2pt,scaled ticks=false,ymajorgrids=true,
grid style={black!12,line width=0.35pt},
label style={font=\small},tick label style={font=\footnotesize},
]
\nextgroupplot[xmin=-0.5,xmax=1.5,ymin=0,ymax=100,xtick={0,1},xticklabels={{Own\\$n=31$},{Unfamiliar\\$n=31$}},xticklabel style={align=center},ytick={0,20,40,60,80,100},ylabel={Mean question score (0--100)},]
\addplot[only marks,mark=*,mark size=2.5pt,draw=white,fill=ownColor,fill opacity=0.82] coordinates {(0.179313289,26.33333333) (0.01708027472,22.16666667) (-0.009742348452,13.83333333) (0.1695120266,50.66666667) (-0.1602859405,40.33333333) (-0.01621671192,40.33333333) (-0.02657822119,32) (-0.06745461102,39.66666667) (-0.1445457732,34.16666667) (-0.1350508767,50.66666667) (0.08360202969,32) (-0.02131951147,51.16666667) (0.1302095408,44.83333333) (-0.1687434128,34.66666667) (0.07444775702,58.16666667) (-0.1448362061,44) (0.1011443023,48.66666667) (-0.008151853894,35.33333333) (-0.03927375122,47) (-0.1307778945,34.16666667) (-0.1714825099,13.83333333) (0.07413809968,43.83333333) (0.04059024658,25.83333333) (-0.04101169994,40.5) (-0.0001356375442,53.83333333) (0.1439556868,46.66666667) (0.02796942886,12.5) (-0.1718751883,50.16666667) (-0.08214154075,33.66666667) (-0.1545697264,30.16666667) (-0.04671582372,37.5)};
\addplot[forget plot,only marks,mark=diamond*,mark size=3.4pt,draw=ownColor,fill=white,line width=0.8pt] coordinates {(0,37.69892473)};
\draw[draw=ownColor,line width=1.2pt] (axis cs:-0.21,39.66666667) -- (axis cs:0.21,39.66666667);
\addplot[only marks,mark=square*,mark size=2.5pt,draw=white,fill=foreignColor,fill opacity=0.82] coordinates {(1.179313289,18.66666667) (1.017080275,3) (0.9902576515,13.66666667) (1.169512027,14) (0.8397140595,13.5) (0.9837832881,4) (0.9734217788,2.166666667) (0.932545389,14.16666667) (0.8554542268,21.66666667) (0.8649491233,13.66666667) (1.08360203,11.16666667) (0.9786804885,21.5) (1.130209541,24.66666667) (0.8312565872,1.666666667) (1.074447757,17.5) (0.8551637939,6.333333333) (1.101144302,20.66666667) (0.9918481461,21.66666667) (0.9607262488,2.666666667) (0.8692221055,23.83333333) (0.8285174901,1.333333333) (1.0741381,11.33333333) (1.040590247,2.5) (0.9589883001,10) (0.9998643625,4.666666667) (1.143955687,10.83333333) (1.027969429,10.83333333) (0.8281248117,7.666666667) (0.9178584592,8) (0.8454302736,3.333333333) (0.9532841763,50.83333333)};
\addplot[forget plot,only marks,mark=diamond*,mark size=3.4pt,draw=foreignColor,fill=white,line width=0.8pt] coordinates {(1,12.62903226)};
\draw[draw=foreignColor,line width=1.2pt] (axis cs:0.79,11.16666667) -- (axis cs:1.21,11.16666667);
\nextgroupplot[xmin=0,xmax=1,ymin=0,ymax=1,xtick={0,0.2,0.4,0.6,0.8,1},ytick={0,0.2,0.4,0.6,0.8,1},xlabel={False-positive rate},ylabel={True-positive rate},]
\addplot[forget plot,black!45,dashed,line width=0.7pt] coordinates {(0,0) (1,1)};
\addplot[forget plot,rocColor,line width=1.2pt] coordinates {(0,0) (0,0.03225806452) (0,0.06451612903) (0,0.09677419355) (0,0.1290322581) (0,0.1612903226) (0,0.1935483871) (0,0.2258064516) (0,0.2580645161) (0,0.2903225806) (0,0.3225806452) (0,0.3548387097) (0,0.3870967742) (0,0.4193548387) (0,0.4838709677) (0,0.5161290323) (0,0.5483870968) (0.03225806452,0.5483870968) (0.03225806452,0.5806451613) (0.03225806452,0.6451612903) (0.09677419355,0.6451612903) (0.09677419355,0.6774193548) (0.09677419355,0.7096774194) (0.09677419355,0.7419354839) (0.09677419355,0.7741935484) (0.09677419355,0.8064516129) (0.09677419355,0.8387096774) (0.09677419355,0.9032258065) (0.1290322581,0.9032258065) (0.1290322581,0.935483871) (0.1290322581,0.9677419355) (0.1612903226,0.9677419355) (0.1935483871,0.9677419355) (0.2258064516,0.9677419355) (0.2903225806,0.9677419355) (0.3225806452,0.9677419355) (0.3870967742,0.9677419355) (0.4193548387,0.9677419355) (0.4516129032,0.9677419355) (0.4838709677,0.9677419355) (0.5161290323,0.9677419355) (0.5806451613,0.9677419355) (0.6129032258,0.9677419355) (0.6451612903,0.9677419355) (0.6774193548,0.9677419355) (0.7096774194,0.9677419355) (0.7419354839,0.9677419355) (0.7741935484,0.9677419355) (0.8064516129,0.9677419355) (0.8387096774,0.9677419355) (0.9032258065,0.9677419355) (0.9032258065,1) (0.935483871,1) (0.9677419355,1) (1,1)};
\node[anchor=south east,font=\small,fill=white,inner sep=3pt] at (axis cs:0.97,0.04) {AUC = 0.934};
\end{groupplot}
\end{tikzpicture}
\Description{A two-panel figure for overall excluding planted error questions. The left panel shows own-paper scores as blue circles in the left column and unfamiliar-paper scores as orange squares in the right column on a 0 to 100 axis. A skipped or timed-out question is scored zero. Each point is one participant's arithmetic mean of the six questions from Unstated rationale, Background knowledge, and Failure mode for that condition. Both observations from a participant use comparable small deterministic horizontal jitter and are not connected. White diamonds mark condition means and colored horizontal segments mark medians. The own condition has n=31, mean 37.7, and median 39.7; the unfamiliar condition has n=31, mean 12.6, and median 11.2. The right panel uses one minus the score proportion to predict the unfamiliar label, which is the positive class; its empirical ROC AUC is 0.934. The diagonal denotes chance-level discrimination. These are descriptive summaries only and do not support inferential or causal conclusions.}
\ifdefined\tikzexternalenable\tikzexternalenable\fi
\endgroup
    \caption{The left plot shows overall score distribution, \textit{excluding planted-error multiple choice questions}, broken down by whether the paper tested is the participant's own paper or the unfamiliar paper ($n=31$ per category). Lines indicate medians and diamonds indicate means. On the right, we plot the ROC for prediction of ``unfamiliar'' label, using $(1-\text{score}/100)$ as the predictor.}
    \label{fig:overall_excluding_pe_score}
    
\end{figure}
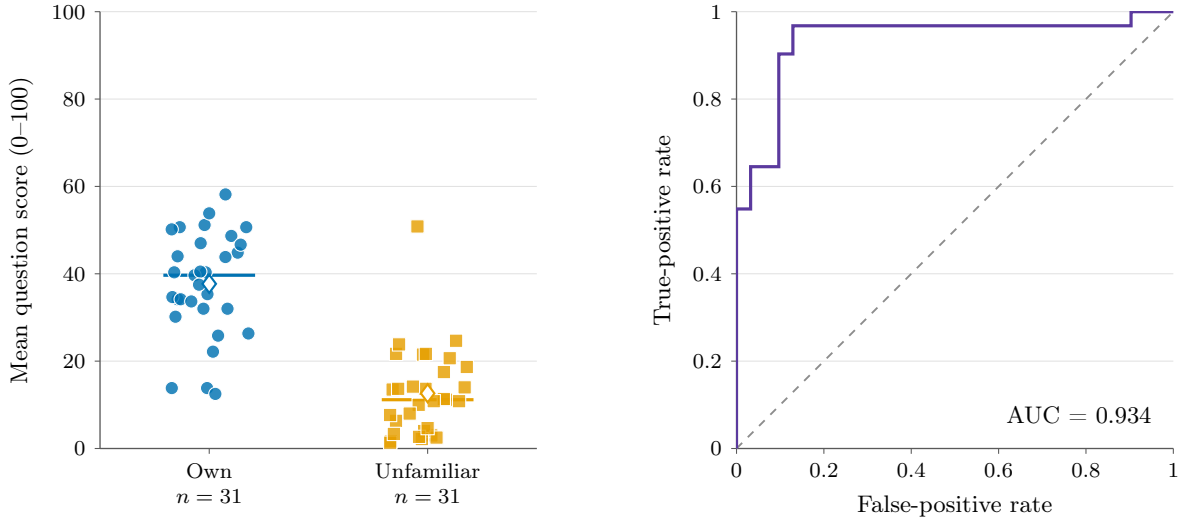

We show the distribution of scores on the participants' own papers and on the unfamiliar papers in~\cref{fig:participant_condition_score_scatterplots}. 
The mean score on participants' own papers was $49.7$, and the mean score on unfamiliar papers was $27.2$. 
On the right side, we compute $(1-\text{score}/100)$ as the class probability for the ``Unfamiliar'' label for all $62$ tests, plot the ROC curve, and report the AUC. 
For this data the AUC turns out to be $0.90$, indicating high discrimination between own-paper and unfamiliar-paper scores.

We also compute the average score on each of the four question families within each test, and plot the distribution of average score within each question type for own and unfamiliar conditions. 
These plots are included in Appendix~\ref{apx:roc_individual}.
We find that the AUC of the planted-error question family (the only multiple-choice family) is very low at $0.595$, which is not surprising, since it primarily measures attention: answers can be found by searching the manuscript directly.
In~\cref{fig:overall_excluding_pe_score}, we see that the AUC increases to $0.934$ when excluding this question family from the score calculation.

These results suggest that our implementation of greCAPTCHA can separate participants who have the capacity to verify from those who do not with a best false-positive rate of around $20\%$.
This rate is lower than typical error rates involved in research assessment. Prior work on the reliability of peer assessment indicates that expert human reviewers can disagree on relative rankings of papers between one-fourth and one-third of the time~\cite{JMLR:v19:17-511}.
Nevertheless, this rate should be minimized as much as possible to minimize unnecessary manual human verification and improve user experiences.
In our qualitative analysis, we show that the main reason for the false-positive rate is the grading rubric rather than question quality.

Supplemental statistical modeling is reported in Appendix~\ref{apx:full-stats}, showing statistically significant differences between scores on all $4$ question families across own and unfamiliar paper conditions.

\subsection{No Evidence of Differentiating Between In-Field and Out-of-Field Papers}
\label{ssec:quantitative:fields}

\begin{figure}
    \centering
    \input{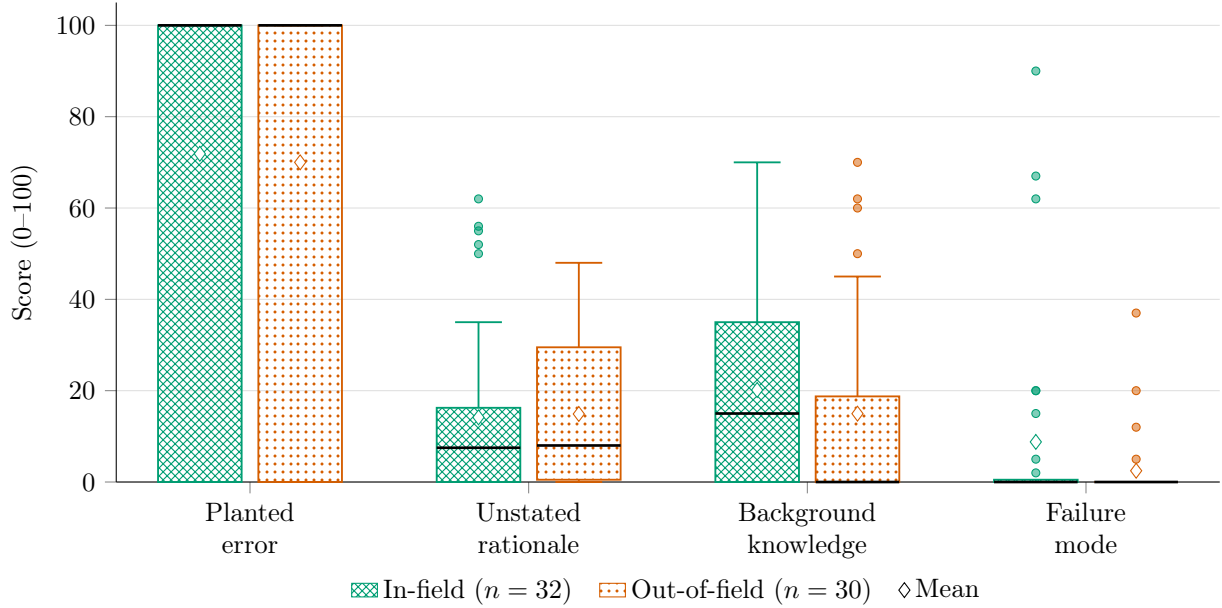}
    \caption{Distribution of unfamiliar-only paper scores grouped by question family
    and field type. Diamonds indicate means. The overall effect of field type is not statistically significant ($p=0.437$).}
    \label{fig:h2_effects}
\end{figure}

To understand the effect of field familiarity on participant performance, we compare the population-level differences between participants in the in-field unfamiliar-paper condition and the out-of-field condition.
The test statistic measuring the difference between in-field and out-of-field scores is estimated to be $0.160$ (average in-field scores are higher) with the corresponding $p$-value estimated at $0.437$. 
We can also see from~\cref{fig:h2_effects} that the score distributions largely overlap between the in-field unfamiliar-paper and out-of-field unfamiliar-paper conditions.
While not definitive proof, this finding does suggest that performance on the question sets generated by our implementation of greCAPTCHA is not dependent on whether the participant is an expert in the field of their assigned unfamiliar paper.
In other words, greCAPTCHA seems to be more than just a ``field detector''.

As explored further in Appendix~\ref{apx:full-stats}, the failure to discriminate between in-field and out-of-field conditions does not appear to be a result of insufficient sample size. A similar analysis shows statistically significant differences between the mean score achieved on the familiar and aggregated unfamiliar papers across all $4$ question families.

One question family where we see a clearer separation between field types is background knowledge.
Here, the scores of in-field participants tend to be higher than those of out-of-field participants.
This finding makes sense intuitively because one would expect domain experts to understand their own field's background better than non-experts.
It also highlights the importance of having multiple question families targeting different types of knowledge.

\section{Qualitative Analysis}
\label{sec:qualitative}

Our qualitative analysis identified $18$ deductive codes such as ``Perceived validity,'' ``Question clarity,'' ``Trust change,'' and ``Feedback relevance.'' 
The full codebook with occurrence frequencies is listed in Appendix~\ref{apx:deductive_codebook}. 
We inductively identified $61$ codes which we organized into $16$ categories.
The complete inductive codebook is provided in~\cref{tab:inductive-codes}.
Here, we interpret our coded excerpts alongside the $31$ participant cases (P1--P31; see Appendix~\ref{apx:participants} for IDs). 

All codes are assigned to individual statements by participants, so it is possible for all deductive and inductive codes to appear multiple times in a single participant's interview transcript. It is also possible for seemingly contradictory codes to be annotated in the same transcript, at different times. For instance, one participant may comment on the strong clarity of one question while also remarking on the poor clarity of another.
Our analysis identified four major themes of feedback from participants regarding:
\begin{itemize}[topsep=4pt]
    \item \textbf{High-level vision (\Cref{subsec:vision}):} Responses about the high-level vision. 
    \item \textbf{greCAPTCHA design choices (\Cref{subsec:concepts}):} Responses to conceptual design choices, such as the four question families or the overall test structure.
    \item \textbf{Implementation details (\Cref{subsec:impl}):} Responses to detailed implementation choices, such as exact question wording, time pressure, or complaints about the UI design. 
    \item \textbf{Deployment (\Cref{subsec:deploy}):} Suggestions for how greCAPTCHA should be deployed. 
\end{itemize}
In each category, we lay out summary points for the positive feedback (+) and negative feedback (--), provide the evidence for each summary point, and give additional context and discussion where necessary.

\subsection{Vision}
\label{subsec:vision}

\paragraph{(+) greCAPTCHA assesses ability to verify.}

Participants mostly agreed that greCAPTCHA is a valid way to assess the capacity to verify. Some highlighted that one would have to go through the actual process of doing the research in order to perform well on the exam (P4, P6, P8, P13, P28, P30). Participant P6 said:

\begin{displayquote}
    \textit{``It would require someone who understands the paper very well, or at least understands the field extremely well, to be able to answer those questions.''}
\end{displayquote}

\paragraph{(+) greCAPTCHA is hard to circumvent.}
A total of 12 participants noted that it would be challenging to prepare for a greCAPTCHA if they had not written the paper.
For instance, P13 reported using AI to walk through the unfamiliar paper beforehand, yet was unable to answer in-depth questions, receiving a score of 1.3 on the unfamiliar paper compared to 51 on their own paper: \simplequote{I just gave it to an AI and asked it to walk me through the whole paper}. Participant P8 commented:
\begin{displayquote}
    \textit{``You needed to know the paper or be able to justify why a certain choice was made in the paper. And as someone who did not read the paper, I could not do that.''}
\end{displayquote}

\paragraph{(+) Enthusiasm for the idea.} 

$23$ participants expressed strong positive overall experiences and enthusiasm for the system, and only $6$ expressed negative experiences.
P28 thought it was important to:
\begin{displayquote}
    \textit{``tackle this problem of people submitting a lot of papers.''}
\end{displayquote}
P7 expressed befuddlement at the prospect of people submitting work for which they do not have the capacity to verify:
\begin{displayquote}
    \textit{``Why would you go to a conference where people present their papers, but then they don't understand. You might as well just stay at home and read the papers.''}
\end{displayquote}

\paragraph{(+) Participants think greCAPTCHA should be deployed.}

$16$ participants expressed some sentiments that greCAPTCHA should be used in deployment, $12$ expressed that it should be deployed after some significant changes, and $7$ expressed sentiments consistent with not deploying. Participants sometimes expressed contradictory sentiments at different times in the interview. For example, P3 expressed initial frustration due to the low evaluation on their own paper, and commented on the confusing nature of some of the questions. Yet when explicitly asked about deployment later in the interview, P3 stated that the motivation is great, and agreed that a future version of greCAPTCHA should be deployed. Participant P12 said:
 \begin{displayquote}
     \textit{``If NeurIPS did it, I think it'd be great.''}
 \end{displayquote}
Participant P18 preferred greCAPTCHA to the version control history solutions currently being employed:
\begin{displayquote}
    \textit{``I honestly would rather see something [like] this as opposed to version control.''}
\end{displayquote}

\paragraph{(+) Deployment would increase participant trust.} 

$22$ participants expressed sentiments consistent with the view that greCAPTCHA's deployment would increase their trust in a peer-review venue. For example, P4 mentioned that deployment:
\begin{displayquote}
    \textit{``would definitely raise my opinion on the quality of work.''}
\end{displayquote}
and P9 said:
\begin{displayquote}
    \textit{``I think it would make me feel a little bit more confident in what I'm reading.''}
\end{displayquote}

\paragraph{(+) Participants guessed the purpose of the system.}

Many participants could guess the purpose of the system, even prior to telling them directly.
$13$ participants guessed the exact purpose of the system and $12$ participants guessed closely, for instance, mentioning the system was intended to measure deep understanding of the paper and/or field. 

\paragraph{(--) Some thought reading the paper or having background knowledge of the field was enough.}

Some participants felt that people may be able to achieve high scores on questions if they had read the paper and/or were familiar with the field.

P13 distinguished understanding a paper from having contributed to it, and P30 provided a counterexample to treating high scores as authorship evidence: after five minutes of preparation, their unfamiliar-paper score was 50.6, close to their own-paper score of 53.1.
They recognized the unfamiliar paper as relevant to their research area, scoring 76.0 on its failure-mode questions versus 45.0 on their own. 
The most senior interviewee by paper count, P12, recalled undergraduate experience while answering an out-of-field medical-physics paper, and scored 55.0 on greCAPTCHA's background-knowledge questions.
Variation in performance may depend on transferable knowledge and long-term experience.

\paragraph{(--) Deployment would require extra work from scientists.}

Although they urged deployment in general, $7$ participants also stressed the importance of maintaining human oversight over decisions. 
This was often expressed as the desire to be able to contest grading decisions, or to have some ranges of scores under which papers would be flagged for human review rather than automatically rejected. However, even this was not unanimous, as P2 stated that they would prefer all decisions to be made autonomously, with results included as ``badges'' for published papers. 

In general, participants emphasized the additional human labor required to maintain human oversight, in addition to the labor of the proctors (P11, P23). However, it is possible that the additional labor required to monitor greCAPTCHA and field complaints would be less than the labor currently expended on full review of large numbers of AI-generated papers.

\subsection{Conceptual Design Choices}
\label{subsec:concepts}

We also asked participants about the conceptual space explored by the questions, and what the grader should assess.

\paragraph{(+) Free-response questions tested deep understanding.}

Overall, $19$ participants
expressed comments that some questions strongly tested deeper concepts, requiring participants to apply, critique, or reconstruct ideas beyond text recall. These questions were usually unstated rationale or failure mode questions.
P26 summarized this sentiment acutely:
\begin{displayquote}
    \textit{``You're reflecting on your own choices and why did you choose path A instead of path B and to really make us ground our decisions in factual data.''}
\end{displayquote}
Participants expressed that questions which test deeper understanding tend to be better at assessing ability to verify. 
Spans of the \code{strongly tested} level of the \code{Question depth} code category overlapped with spans of the \code{agree} level of the \code{Perceived validity} code category $9$ times in $5$ interviews.

\paragraph{(+) Opportunity for learning.}

The value of greCAPTCHA extended beyond assessment.
P25 remarked of a question, \pquote{I'm shocked in the number of times I've presented this research that no one has asked me this question,}{P25} and indicated that it would be useful to think about further.
P20 acknowledged that the system found genuine gaps in their understanding and requested we send them their greCAPTCHA questions for further study: \pquote{Can I get a copy of these questions?}{P20}. 
P26 similarly interpreted the encounter as a lesson to understand why they made particular decisions.
Low scores did not preclude these reactions: P20 had the lowest own-paper score, and P26 scored 34.4. 

Some participants even proposed changing the purpose of the system.
P15 articulated how their attitude would change accordingly: \simplequote{I think if it was teaching me, I would be much more responsive to it.}
\begin{displayquote}
    \textit{``If you would generate a paper, I guess you have to learn what's in the paper and then try to learn about the literature that it's drawn from.''}---P7, on how to prepare for a greCAPTCHA.
\end{displayquote}

\paragraph{(+ and --) Some participants performed poorly on some/all questions for their own papers.} 

Some participants scored poorly on background knowledge questions in their own papers.
P9 acknowledged not fully understanding statistics initially supported by a mentor and recognized that some gaps remained. 
P26 described following established statistical practices without fully understanding their rationale: \pquote{We just know we have to use this statistical stuff for our results. So we do it, we don't really understand super why sometimes.}{P26}. 
Their own-paper background-knowledge scores were 0.0 (P9) and 17.5 (P26). 
These results appear to be a correct assessment of inability to verify concepts. As mentioned above, some participants took their low scores as an opportunity for learning.

In contrast, P3 experienced a sharply different meaning of their own low score (22.9). 
Confronted with a rubric that rejected their account of something they had devised, they protested, \pquote{I made this up myself, I made this myself!}{P3}. 
P10's difficulty with a closely related unfamiliar paper resulted in uncertainty about their own competence: \pquote{Am I misunderstanding my field? Am I misunderstanding the papers?}{P10}. 
Their unfamiliar-paper score was 8.4, with 30 minutes of preparation, and they even suspected that the unfamiliar paper had significant flaws but then hesitated to trust their judgment. 
This encounter illustrates how an apparently authoritative assessment can make a researcher doubt the expertise needed to challenge it.

Three participants scored worse on their own papers than the unfamiliar papers (P1, P3, P20). These were junior participants. This probably indicates a true lack of understanding of their fields (+), but we should be careful about punishing junior authors too harshly for this. We see some evidence that authoritative and negative AI-generated assessments may be discouraging for junior authors (--).

\paragraph{(--) Some questions were considered irrelevant.}

P6 described an algorithm question that essentially only they could answer, yet called it a detail that \pquote{doesn't really matter for any of the big conclusions in the paper}{P6}. 
Other participants also felt that questions were not relevant (mentioned by $16$ participants in $41$ instances). P29 noted \pquote{[The system] has the tendency of going to technical details and picking out some detail and then ask you just another detail, but miss the bigger picture}{P29}.

Participants were not told the purpose of the test was to assess capacity to verify. Questions which asked about specific details may help to discriminate authorship or ability to verify without being perceived as ``relevant'' to the author. This trade-off between relevance and assessment capacity is important to consider during deployment, but it does not necessarily detract from the test's primary purpose of assessing understanding.

\paragraph{(--) Multiple-choice questions are too easy.}

When participants remarked on questions that were too easy, they often mentioned the ability to \textit{``control-F''} (that is, search) terms in the question (P5, P8, P9, P11), specifically citing the multiple-choice questions (P14).  
This intuition corroborates the quantitative analysis showing that the multiple-choice questions are not discriminative and even \emph{harm} predictiveness of the evaluation.
On the basis of both the quantitative and qualitative results, we recommend removing the multiple-choice questions in future iterations of greCAPTCHA.

\paragraph{(--) Need to clarify grading expectations.}

Participants' comments on the grader were generally negative.

Overall, $23$ participants described uncertainty about the expected level of detail in responses. 
P14 imagined a conversation at a conference poster session, while P6 explained concepts assuming little technical background.
P16 gave a succinct summary of the core reason for this variance: 
\begin{displayquote}
    \textit{``What audience am I explaining these answers to? [...] The way that I would explain it to a student who is in their first course, or am I explaining it to a professor who is talking about this specific nitty-gritty paper?''}---P16, on being unclear who is the expected recipient of their answer.
\end{displayquote}
Participants' individual choices affected what they made explicit.
P11 objected to having to guess the expected answer or restate implications they considered obvious. 
P25 offered a counterpoint: \simplequote{the text [manuscript] is all the grader has}, suggesting that from an examiner's perspective, omitted reasoning cannot automatically be inferred or credited.
This is an important gap that concerns the evidence required to demonstrate the capacity to verify: disciplinary conventions may be meaningful to a colleague but are likely to remain insufficient for the grading of a rubric.

Overall, unclear expectations led to $16$ participants disagreeing with some aspect of the grading, while only $8$ expressed agreement with the grading. 
Several participants focused on the very strict allegiance to the rubric (P9, P13, P20, P21).
P21 summed this up by saying: \simplequote{I think if I would imagine a human reading this compared to the tool, I would picture that the human is more charitable.}
On the other hand, P8 found the system provided explanations helpful, and P25 even accepted some deductions.

The feedback on the grading suggests several improvements to greCAPTCHA:
\begin{itemize}[topsep=4pt]
    \item The expected level of detail should be stated clearly at the beginning of the exam, and the questions and rubrics should be generated using this statement.
    \item If something is included in the grading rubric, it should be clearly signaled in the question text. The paper's author should be able to respond to the questions exactly as phrased and get 100\%.
    \item The grading rubric should be more flexible and award points for answers that are not precisely what was initially expected but are nevertheless correct.
    \item It may also be possible to assess answers not by their correctness to the letter of the questions, but by the likelihood that the examinee understands the paper. An examinee may misunderstand the question text, but still demonstrate strong understanding of the paper in an ostensibly incorrect answer.
\end{itemize}

\subsection{Implementation Details}
\label{subsec:impl}

We discuss responses to low-level implementation decisions, such as the UI design or exact question phrasing.

\paragraph{(+ and --) Time pressure.} 

Time pressure also appeared in $20$ participants' coded accounts. 
Its consequences depended on both familiarity and strategy. 
P20 reported spending half the available time on the first own-paper question.
Overall, we recorded three own-paper timeouts at an average score of 10.4. 
P1 described how time pressure restricted them from elaborating familiar material and led to a consequential typo.
P15 mistakenly believed skipped questions could be revisited, although the interface clearly stated that skipping a question would result in a grade of $0$ on that question.
Their unfamiliar-paper attempt contained six skips, resulting in a score of 17.3. 
P15 also reported profound unfamiliarity, so the low score cannot be attributed solely to the interface.
Such cases show how knowledge, pacing, and interface interpretation jointly shape the recorded response. These factors made some participants perform poorly on their own paper and others perform poorly on the unfamiliar paper.

Our time was limited during the study, and we could afford participants only 15 minutes per paper. However, an actual deployment may allow examinees more time to answer the questions.

\paragraph{(--) Question wording was often unclear.}

Participants also often felt that question wording and the expected response were unclear.
$21$ participants discussed this topic.
For example, P18 mentioned that for one question, \simplequote{it's hard to verbalize whether or not you understand the message or even relay: `yes, I know what this detail means', because it's buried in the question.} 
P18 discussed how some questions had multiple parts that were not organically connected to each other, making them difficult to answer adequately. As a result, they had to take extra time to think deeply about their response, which can be challenging during a timed exam.

LLMs should be prompted to ask about only one aspect per question and to avoid lengthy sentences with complex structures.

\paragraph{(--) Three participants got questions they couldn't answer but co-authors could.}

Question specificity may also create difficulties for collaborative authorship. 
P5 distinguished their theoretical knowledge from a co-author's practical implementation decisions; P19 described how they contributed an implementation without knowing the reasoning behind some metrics; and P29 encountered material written by a co-author in the greCAPTCHA. 
Their own-paper scores were 55.3, 50.6, and 47.6, respectively. P5's response was \simplequote{I should go ask my co-author why he did it this way}.
These accounts highlight the importance of accurately matching participants' stated contributions with greCAPTCHA's questions.
We may wish to change the way we ask about contributions to make it clearer that you will be tested on what you state you can be tested about, so you should indicate exactly what technical elements your co-authors contributed that you don't understand. While our participants were blinded to the purpose of the study, knowledge of its purpose may result in participants describing their contributions more clearly in the initial submission form. Finally, in the test itself, we could make it possible to excuse oneself from certain concepts during the exam, for instance by responding ``I didn't work on this part'' to a question. We can give examinees a limited budget to do this. 

\subsection{Deployment}
\label{subsec:deploy}

As discussed in \Cref{subsec:vision}, participants largely agreed with deploying a future iteration of greCAPTCHA. However, we heard a wide range of opinions about how exactly to implement and deploy greCAPTCHA in practice. 

Participants had contrasting visions for how greCAPTCHA reports should be used in real workflows.
Some participants expressed support for entirely automating some decisions (P2, P7, P8, P16), especially in the most egregious cases. 
P2 preferred converting AI-assigned scores into badges for papers, based on defined score thresholds. 
P26 noted that setting an acceptable threshold would be challenging. 
Five participants suggested that there should be three categories of papers: those that are automatically desk rejected, those that require additional screening, and those that are automatically passed into the primary reviewing pool (P7, P13, P14, P15, P20).
Several participants preferred maintaining an appropriate level of human oversight before making decisions (P7, P14, P26, P28).
P14 suggested that humans could edit questions and rubrics prior to assessment, and also suggested that greCAPTCHA scores be used for metareview.
P18 emphasized that \simplequote{to implement a tool, there would have to be some good faith process to contest}.
Other participants noted that simply implementing the assessment may have a deterrent effect, either due to the extra work involved or due to the cognitive effects of receiving a low score (P3). 
\begin{displayquote}
    \textit{``If I had to choose between a conference that allowed me to just submit a paper and [...] a conference that required me to take a half an hour test on my own paper, I would have to really be invested in that conference to do so, which might be a good thing because there's oversubmission problems at conferences.''}---P18, on the deterrent effect of greCAPTCHA.
\end{displayquote}
These proposals frame disclosure itself as a consequential design decision.
P18's call for a good-faith contest process and P28's preference for human evaluation place responsibility for resolving uncertainty with the institution, although P24's comments color this approach: \simplequote{it is extra work for people who care, who try to do ethical research. But someone who doesn't care [...] that [greCAPTCHA] is still not going to stop them.}
Several participants suggested that reviewers be able to see the system report (P16, P20, P21, P27, P28).
However, other participants felt strongly that \pquote{flagging it for reviewers is not helpful because then you unnecessarily bias your reviewers}{P11}.

Other participants suggested modifying the modality of the assessment ($3$ participants) or allowing back-and-forth interactions ($7$ participants).
They frequently cited a desire to clarify their answers or push back on the assessment. 
They also discussed a desire to request clarification on question texts during assessment. These concerns would be easier to address in an audio or video format.

Finally, some participants emphasized the potential unintended measurements beyond the capacity to verify.
These concern how knowledge is elicited under examination conditions.
Three participants raised issues with language use and reading speed (P4, P18, P21) regarding accessibility for non-English speakers and people with disabilities.
P4 summarized the distinction: \simplequote{just because I can't write it in the same way that the question is expecting doesn't mean that I don't know about the work.} 
Typing-related remarks appeared in 12 participants' coded accounts.
P19 described the task as testing \simplequote{both my knowledge of the paper and also my ability to info-dump about it.} 
Relatedly, three participants explicitly linked performance to how well people can search for information in a document (P7, P18, P27).
P18 described the effect on performance of \simplequote{how efficiently they can use the search function in a PDF reader}, while P8 offered a more favorable interpretation: knowing where to find an answer quickly can itself reflect familiarity.
Recency may also play a role.
P1 associated difficulty with having left their research area for industry, and P19 mentioned that two years had passed between finishing their project and its acceptance, during which intervening projects made decisions harder to recall.
P12 provided a counterexample: \simplequote{I'm giving myself a lot of free credit for a paper I haven't read in 12 years}, yet their own-paper score was 58.6, compared with 32.3 for P1 and 50.6 for P19.


\section{Conclusions} 

In this paper, we introduced greCAPTCHA, an assessment method designed to measure an author's understanding of a research manuscript. We propose that this assessment serve as evidence of research authorship. We measure this evidence via the construct of ``capacity to verify,'' which we define as the necessary knowledge and reasoning skills required to critically evaluate the contributions for which authors claim responsibility. We developed a prototype of greCAPTCHA and evaluated its usefulness via an in-person user study and semi-structured interviews with 31 participants. Using GPT-5.6 Sol, greCAPTCHA generates tailored questions to assess each author's distinct contributions to their manuscript, using an evaluation tailored directly to their work. These questions are designed to measure the authors' understanding of a manuscript at four distinct levels: planted errors, unstated rationales, background knowledge, and failure modes.

To evaluate greCAPTCHA, we used qualitative and quantitative measures to assess participants across two scenarios: an unfamiliar paper selected by the research team according to a standardized procedure, and a paper they had authored, with the latter assessment tailored to their self-specified contributions. Our findings show that participants performed significantly better on their own work than on the unfamiliar control papers (with an AUC of 0.90). Our analysis showed no statistically significant overall difference in performance on unfamiliar papers inside or outside a participant's field. This suggests that greCAPTCHA successfully isolates manuscript-specific understanding from general domain familiarity. More specifically, these results demonstrate that our prototype can differentiate between participants who have the capacity to verify and those who do not, achieving a lowest false-positive rate of 20\%. We believe that this error rate should be further reduced, and we discuss several directions for doing so. Participants generally viewed greCAPTCHA as a plausible and robust method for assessing manuscript understanding and therefore as evidence of research authorship.

Participants indicated that deploying a well-designed greCAPTCHA would increase their trust in peer review as a verification mechanism. They also identified free-response questions on rationales and failure modes as particularly effective in evaluating deep understanding of their authored manuscript. Conversely, participants noted that ambiguous question wording, unclear grading expectations, overly rigid rubrics, time constraints, and questions extending beyond an author's specified contribution could result in misleadingly low scores. This feedback points to a clear need for rigorous greCAPTCHA design standards to ensure the tool remains both fair and diagnostic.

As an assessment method, greCAPTCHA shifts the focus of authorship attribution from inspecting the final text artifact (the model of tools like Pangram) to validating the intellectual process behind it. By anchoring verification to the author's capacity to understand and critique their manuscript, our proposed method motivates authentic engagement with the research. Translating this approach to real-world environments requires clear governance in which proposed questions to authors must clearly match declared contributions, evaluation rubrics must remain transparent and adequate, and human oversight must govern consequential judgments. Although our study established the efficacy of greCAPTCHA within a controlled, in-person testing environment, the method holds broad promise for academic publishing and university pedagogy alike. We suggest that conferences, journals, educators, and funders begin prototyping localized implementations. From the lessons learned, greCAPTCHA could evolve into a standardized mechanism that accommodates responsible AI tools while ensuring authors can fully stand behind the claims they publish and receive appropriate credit.

\section*{Acknowledgments}
We thank our anonymous study participants, as well as Vishisht Rao for piloting the study.
B.G.'s work was funded in part by the Institute for Complex Social Dynamics at Carnegie Mellon University.
A.K.'s work was funded in part by the AI2050 program at Schmidt Sciences (grant 24-66924).
N.S., J.P., and B.G. were funded in part by grants NSF 1942124 and ONR N000142512346. 
This work was also supported in part by the Alfred P. Sloan Foundation under Grant No. G-2026-79567. 
We thank the Gemini Academic Program Award for credits to use Gemini.

\printbibliography

\appendix
\definecolor{promptbg}{gray}{0.97}
\definecolor{promptframe}{gray}{0.75}
\definecolor{promptnum}{gray}{0.55}

\lstdefinestyle{promptfile}{
  basicstyle=\ttfamily\scriptsize,
  backgroundcolor=\color{promptbg},
  frame=single,
  rulecolor=\color{promptframe},
  framesep=4pt,
  xleftmargin=2.5em,
  xrightmargin=0.5em,
  numbers=left,
  numberstyle=\ttfamily\tiny\color{promptnum},
  numbersep=8pt,
  stepnumber=1,
  breaklines=true,
  breakatwhitespace=false,
  breakindent=0pt,
  postbreak=\mbox{\textcolor{promptnum}{$\hookrightarrow$}\space},
  columns=fullflexible,
  keepspaces=true,
  showstringspaces=false,
  tabsize=2,
  upquote=true,
  extendedchars=true,
  literate={—}{{\textemdash}}1,
}%

\section{Question Generation Prompts}
\label{apx:question-prompts}

This section contains~\cref{lst:planted-error,lst:unstated-rationale,lst:background-knowledge,lst:failure-mode} which each contain the verbatim prompts used with greCAPTCHA to generate the questions sets and grading rubrics of our experiment.

\begin{lstlisting}[style=promptfile,caption={The greCAPTCHA prompt used for generating \textbf{planted-error detection} questions.},label={lst:planted-error}]
Generate planted-error detection items. Each item states a specific claim about this manuscript, and the participant must identify which version of the claim is what the paper actually reports.

Build each item from a single atomic, verifiable fact that appears exactly once in the paper: a numeric result, an ablation delta, a dataset or baseline name, a hyperparameter, a section or table attribution. Anchor the item to where that fact lives, such as "the ablation in Section 5.2" or "Table 3", so the key is checkable against a specific passage. Never include Section or Table names in the question text to avoid easily looking up the information.

Perturb a value or a referent, never a qualitative direction. Reassigning which component an effect belongs to, or which condition a number describes, is the target. Flipping "improves" to "degrades" is too easy and must not be used. Every incorrect option must be plausible enough that a reader who does not know the work has to locate and read the relevant passage to rule it out.

Never build an item on a fact the paper restates elsewhere, including in the abstract, a figure caption, or the conclusion. A restatement gives a non-author a cheap second place to check. Never build an item whose correct option can be identified by keyword overlap with the prompt, by grammar, by option length, or by being the most specific or most hedged option. Options must be mutually exclusive and comparable in length, specificity, and technical register.

Discard any candidate answerable from the title and abstract alone, and any candidate that someone who knows this field but has never read this paper would get right.
\end{lstlisting}

\begin{lstlisting}[style=promptfile,caption={The greCAPTCHA prompt used for generating \textbf{unstated rationale} questions.},label={lst:unstated-rationale}]
Generate unstated-rationale items. Each item asks the participant to justify a methodological or design choice whose reason is not stated in the manuscript.

Locate a choice the paper makes but does not explain: a test or estimator selected over an obvious alternative, a threshold or cut-off, an excluded condition, an ordering of stages, a control that is present or conspicuously absent. Ask why that choice is appropriate here and, where it sharpens the item, why it would not be appropriate for a specific other part of the paper.

The reason must be genuinely absent from the manuscript, so that no amount of searching the PDF produces it. If the paper explains the choice anywhere, including a footnote, an appendix, or a limitations paragraph, discard the item.

Ground every item in a specific named element of this paper: this comparison, this table, this preprocessing step. A question that could be asked of any paper in the field is testing field knowledge rather than authorship of this artifact, and must be discarded.

Build the rubric to award credit for reasoning that connects the choice to the specific properties of this study's data or design, and to withhold credit for a generically correct textbook justification that never touches this paper. Accept substantively equivalent reasoning and multiple valid explanations. Award no credit for fluency, length, confidence, or command of English; score only the content of the reasoning.
\end{lstlisting}

\begin{lstlisting}[style=promptfile,caption={The greCAPTCHA prompt used for generating \textbf{background knowledge} questions.},label={lst:background-knowledge}]
Generate background-concept items. Each item asks the participant to define or explain, in their own words, one concept the paper depends on but does not itself define — the presupposed background a competent member of this field carries into reading it, not the paper's own contribution.

Choose concepts that are load-bearing. If the participant misunderstood the concept, the paper's design, its choice of method, or its interpretation of its results would no longer make sense. A term mentioned once in passing is not load-bearing; the standard is that you could name a specific design decision or claim that depends on it.

The most important filter is that the paper must not define the concept. The participant answers with the paper in front of them, so a concept the paper explains in its own words is a lookup rather than a question. Prefer concepts the paper names and uses as though they need no introduction. Where the paper's own topic is a concept it does define, choose an adjacent concept it presupposes instead: for a paper about p-hacking that defines p-hacking, ask what a p-value means under the null hypothesis, what the familywise error rate is, or what pre-registration is for.

Never name a section, table, figure, or page in the question text, and never reuse the paper's own phrasing of the concept. Both point the participant at a passage to copy.

Ask for two things in each item, in this order: the definition or explanation itself, and one sentence on why the concept matters for the work reported here. The first half is what the item is for. The second half is what someone who does not understand the work cannot supply, and it is where an answer assembled from general knowledge alone reads as generic.

Build the rubric as an enumeration of the elements a correct definition must contain, each its own criterion with its own points, totalling 70, plus one criterion worth 30 for the connection to this paper. Name the required elements explicitly in the guidance, so grading is a check against a list rather than a judgement of quality. For at least one criterion, state a specific plausible-but-wrong answer that earns nothing — the near-miss someone with topic-adjacent familiarity would give — so an answer that circles the concept without stating it cannot collect points.

Award nothing for fluency, length, hedging, confidence, or restating the question. Accept any wording that carries the required elements, including informal phrasing, an example that entails the definition, and notation in place of prose. Do not require the participant's terminology to match the paper's.

Discard any candidate concept whose definition appears anywhere in the paper, any answerable from the title alone, and any that duplicates a concept an earlier card already covers.
\end{lstlisting}

\begin{lstlisting}[style=promptfile,caption={The greCAPTCHA prompt used for generating \textbf{failure mode} questions.},label={lst:failure-mode}]
Generate one item asking the participant to name a realistic condition under which this work's method or central finding would degrade, and to say why it would.

Keep the question to one or two sentences, and ask for a short answer of two or three sentences. Word it neutrally: a condition the work was not built for, not a flaw in it. A scored item that reads as an attack on the participant's own paper invites a defensive answer rather than an informative one.

The condition must be specific to this setup and plausible in this domain — a property of the data, a regime, a scale, a population, or an interaction this pipeline would mishandle. Its mechanism must follow from how this work is built, not from general methodological caution.

Do not use anything the paper names itself: its limitations, its future work, its statements about what it did not test, or anything in the abstract. The participant reads with the paper open, so those are lookups. Never name a section, table, or figure in the question text.

In the guidance, list the qualifying conditions for this paper, each with the mechanism that makes it fail. Any one of them earns the naming points, since reasonable people will pick different edges.

Then list the answers that earn nothing: that the sample is small, that the results may not generalise, that more data or more baselines are needed, that the method is untested in other settings, and anything else that could be written without having read this paper. A fluent, confident answer of exactly that kind is the most likely wrong answer here, and it must score zero.

Weight the mechanism above the condition. Award nothing for fluency, length, hedging, or for restating what the paper already says about its own scope.
\end{lstlisting}

\newpage
\section{Paper Selection Prompt Template}
\label{apx:paper-selection-prompt}

The prompt used to suggest five potential candidates for the unfamiliar paper of the participant, shown in~\cref{lst:paper-selection-prompt}.
This prompt generates a full report with justification of why each of the five candidate papers are appropriate.

\begin{lstlisting}[style=promptfile,caption={The prompt used to select five candidate unfamiliar papers for a participant.},label={lst:paper-selection-prompt}]
You are helping select a paper for a controlled study. Attached is a manuscript written by {{PARTICIPANT_NAME}}, one of its authors. I need five candidate papers by other people that this person is unlikely to have read, all of them {{IN-FIELD/OUT-OF-FIELD}} for them.

Definitions, which govern everything below.

IN-FIELD means their training transfers: the notation, methods and background concepts are ones they already hold, and they could read the paper without learning a new apparatus. Same research area, different research group.

OUT-OF-FIELD means their training does not transfer: the background concepts are ones they have no reason to know. It does not mean unreadable. The paper must still be a normal empirical or theoretical paper that an intelligent outsider can follow at the level of its claims, because a participant who cannot parse the abstract at all tells us nothing.

STEP 1 — Identify the person, and say how.

Use the attached manuscript as ground truth: its author list, affiliation, topic and references are the anchor. Then search for this person's publication record.

Report what you found as a short profile: their research area, subfield, the methods they work with, their affiliation and career stage, their most cited work, and the venues they publish in. Cite a source for each claim — their scholar profile, lab page, arXiv listing, or a specific paper.

Name disambiguation is the failure mode that ruins everything downstream. If more than one researcher shares this name, say so explicitly, state which one you concluded is the author of the attached manuscript, and give the evidence that settles it. If you cannot settle it confidently, stop and tell me rather than guessing.

STEP 2 — Define their familiarity neighbourhood.

Before choosing anything, list what this person is likely to have already read:
- their own papers and their coauthors' papers;
- work by their advisor, their lab, and their institution's group in this area;
- the papers their attached manuscript cites, and well-known papers that cite it;
- the canonical or widely discussed papers of their subfield, including anything that circulated heavily on social media or won an award;
- the standard venues they attend, whose proceedings they will have browsed.

Everything in that neighbourhood is disqualified. Say what you excluded and why, briefly.

STEP 3 — Select five candidates.

Constraints on every candidate:
- A real paper, posted to arXiv in 2025 or 2026, with a resolvable arXiv ID. Quote the first sentence of its abstract verbatim as evidence you actually retrieved it. Do not offer a paper you cannot verify exists; five verified candidates matter more than five interesting titles.
- Not authored by this person, their coauthors, their advisor, or their institution.
- Outside the familiarity neighbourhood from Step 2.
- Low visibility: few citations, no award, no viral thread. A paper they would have encountered by accident is useless to me.
- Comparable in difficulty to the attached manuscript — similar length, similar density of mathematics, a similar number of experiments or results. If the unfamiliar paper is markedly harder than their own, any score difference I measure is about difficulty rather than about authorship, which destroys the comparison.
- Self-contained enough to be read cold: it states its own research question and what it found, rather than depending on a companion paper.

Make the five diverse, so that one wrong guess about this person does not invalidate the whole set. For in-field, five different research groups. For out-of-field, five different fields.

STEP 4 — Report.

Rank them best fit first, and for each give:
1. Title, authors, arXiv ID and link, posting date, primary arXiv category.
2. The first sentence of the abstract, quoted.
3. One sentence on what the paper is about.
4. Why it is {{IN-FIELD/OUT-OF-FIELD}} for this person specifically, referring to the profile from Step 1 rather than to the field in general.
5. Why they are unlikely to know it, referring to the neighbourhood from Step 2.
6. Difficulty match to the attached manuscript: length, mathematical density, number of results, and whether you judge it easier, comparable, or harder.
7. The single most likely reason this person might in fact already know it. Every candidate gets one; if you cannot think of one you have not looked hard enough.
8. Your confidence that they have not read it, as high, medium or low.

End with the one candidate you would pick if you had to choose, and the one you consider riskiest, each in a sentence.

Do not pad the list to five with papers you doubt. If you can only verify three that meet the constraints, give me three and say what blocked the others.

\end{lstlisting}

\clearpage
\section{Participant Summary}
\label{apx:participants}

\begin{table*}[h]
  \centering
  \caption{Participant characteristics and metadata for participants' own and unfamiliar papers. Preparation time is the amount of time the participant reported reviewing the unfamiliar paper, in minutes.}
  \Description{A table with one row for each of 31 participants. Participant-background columns report identifier, career position, publication-count range, and research field. Own-paper columns report publication year and field. Unfamiliar-paper columns report field, page-count range, and preparation time in minutes. The participants span master's students, PhD students, postdocs, and one professor, and their unfamiliar papers cover both matching and nonmatching research fields.}
  \label{tab:participants}
  \small
  \setlength{\tabcolsep}{3pt}
  \renewcommand{\arraystretch}{1.12}
  \begin{tabularx}{\textwidth}{@{}rlr>{\RaggedRight\arraybackslash}Xr>{\RaggedRight\arraybackslash}Xrr@{}}
    \toprule
    & \multicolumn{2}{c}{Participant background}
    & \multicolumn{2}{c}{Own paper}
    & \multicolumn{2}{c}{Unfamiliar paper}
    & \\
    \cmidrule(lr){2-3}\cmidrule(lr){4-5}\cmidrule(lr){6-7}
    \textbf{ID} & \textbf{Position} & \textbf{Pubs.}
    & \textbf{Field} & \textbf{Year}
    & \textbf{Field} & \textbf{Pages}
    & \shortstack[r]{\textbf{Prep. time}} \\
    \midrule
    1 & Master's & 0--4 & Machine Learning & 2026 & Astrophysics & 1--10 & 60 \\
    2 & Postdoc & 5--9 & Social and Information Networks & 2026 & Human AI Interaction & 11--20 & 0 \\
    3 & Master's & 0--4 & Logic & 2026 & Geophysics & 11--20 & 0.5 \\
    4 & PhD & 0--4 & Robotics & 2026 & Geophysics & 11--20 & 0 \\
    5 & PhD & 5--9 & Logic in Computer Science & 2026 & Quantitative Biology & 11--20 & 30 \\
    6 & PhD & 0--4 & Programming Languages & 2026 & Chemical Physics & 21--30 & 5 \\
    7 & PhD & 0--4 & Computer Science and Game Theory & 2024 & Artificial Intelligence & 11--20 & 5 \\
    8 & PhD & 5--9 & Geophysics & 2026 & Image and Video Processing & 11--20 & 10 \\
    9 & PhD & 0--4 & Human-Computer Interaction & 2026 & Education Technology & 1--10 & 6 \\
    10 & Postdoc & 5--9 & Signal Processing & 2024 & Computer Vision and Pattern Recognition & 11--20 & 30 \\
    11 & PhD & 5--9 & Distributed, Parallel, and Cluster Computing & 2026 & Theoretical Computer Science & 21--30 & 20 \\
    12 & Professor & 50+ & Neurons and Cognition & 2015 & Medical Physics & 21--30 & 0 \\
    13 & Master's & 0--4 & Computation and Language & 2026 & Computational Linguistics & 11--20 & 10 \\
    14 & PhD & 5--9 & Artificial Intelligence & 2026 & LLMs/Agents & 21--30 & 5 \\
    15 & Postdoc & 20--49 & Human-Computer Interaction & 2021 & Astrophysics & 11--20 & 0 \\
    16 & Postdoc & 5--9 & Computation and Language & 2025 & Computation and Language & 21--30 & 25 \\
    17 & Postdoc & 10--19 & Geophysics & 2025 & Robotics & 1--10 & 30 \\
    18 & Postdoc & 5--9 & Neurons and Cognition & 2026 & Atmospheric and Oceanic Physics & 41--50 & 0 \\
    19 & PhD & 0--4 & Robotics & 2025 & Robotics & 1--10 & 5 \\
    20 & PhD & 0--4 & Machine Learning & 2026 & Fluid Mechanics & 21--30 & 0.5 \\
    21 & Postdoc & 5--9 & Human-Computer Interaction & 2026 & Human-Computer Interaction & 11--20 & 20 \\
    22 & PhD & 5--9 & Machine Learning & 2026 & Clinical AI & 21--30 & 0 \\
    23 & PhD & 5--9 & Human-Computer Interaction & 2022 & Neurons and Cognition & 11--20 & 0 \\
    24 & PhD & 5--9 & Cryptography and Security & 2025 & Populations and Evolution & 21--30 & 20 \\
    25 & Postdoc & 5--9 & Neurons and Cognition & 2026 & (Computational) Linguistics & 21--30 & 10 \\
    26 & PhD & 0--4 & Logic in Computer Science & 2026 & Biological Physics & 11--20 & 5 \\
    27 & PhD & 0--4 & Machine Learning & 2026 & Soft Condensed Matter & 11--20 & 0 \\
    28 & PhD & 10--19 & Cryptography and Security & 2026 & Microfluidics & 11--20 & 5 \\
    29 & Master's & 0--4 & Neurons and Cognition & 2026 & Logic in Computer Science & 11--20 & 0 \\
    30 & PhD & 5--9 & Computers and Society & 2026 & Education Technology & 11--20 & 5 \\
    31 & PhD & 0--4 & Geophysics & 2024 & Geophysics & 51--60 & 3 \\
    \bottomrule
  \end{tabularx}
\end{table*}

\clearpage
\section{Distribution of Scores by Question Family}
\label{apx:roc_individual}

\begin{figure}[h]
    \centering
\begingroup
\ifdefined\tikzexternaldisable\tikzexternaldisable\fi
\definecolor{ownColor}{HTML}{0072B2}
\definecolor{foreignColor}{HTML}{E69F00}
\definecolor{rocColor}{HTML}{5B3A9E}
\begin{tikzpicture}
\begin{groupplot}[
group style={group size=2 by 1,horizontal sep=0.16\linewidth},
scale only axis,
width=0.35\linewidth,height=0.35\linewidth,
axis lines*=left,axis line style={black!65,line width=0.5pt},
tick align=outside,tick style={black!65,line width=0.4pt},
major tick length=2pt,scaled ticks=false,ymajorgrids=true,
grid style={black!12,line width=0.35pt},
label style={font=\small},tick label style={font=\footnotesize},
]
\nextgroupplot[xmin=-0.5,xmax=1.5,ymin=0,ymax=100,xtick={0,1},xticklabels={{Own\\$n=31$},{Unfamiliar\\$n=31$}},xticklabel style={align=center},ytick={0,20,40,60,80,100},ylabel={Mean question score (0--100)},]
\addplot[only marks,mark=*,mark size=2.5pt,draw=white,fill=ownColor,fill opacity=0.82] coordinates {(-0.05683042952,50) (0.006422494196,100) (-0.05795936921,50) (-0.1736223426,50) (0.06650379513,100) (0.1449664001,100) (0.07923472103,50) (-0.06564199459,100) (0.02905047486,100) (-0.1017440511,50) (-0.03266923943,50) (0.07306865243,100) (-0.1008320982,100) (-0.06823804907,100) (-0.1207563177,100) (0.01234594042,100) (0.02737931212,100) (-0.07464393478,100) (0.03721143451,100) (-0.1190560139,100) (0.08188846361,0) (-0.0360587841,100) (-0.1215745642,100) (0.1350690564,100) (0.1140006794,50) (-0.1703760352,100) (0.1704116568,100) (0.0873860417,100) (-0.01115004531,100) (0.07803263561,100) (-0.1123715738,100)};
\addplot[forget plot,only marks,mark=diamond*,mark size=3.4pt,draw=ownColor,fill=white,line width=0.8pt] coordinates {(0,85.48387097)};
\draw[draw=ownColor,line width=1.2pt] (axis cs:-0.21,100) -- (axis cs:0.21,100);
\addplot[only marks,mark=square*,mark size=2.5pt,draw=white,fill=foreignColor,fill opacity=0.82] coordinates {(0.9431695705,100) (1.006422494,100) (0.9420406308,100) (0.8263776574,100) (1.066503795,100) (1.1449664,100) (1.079234721,50) (0.9343580054,100) (1.029050475,100) (0.8982559489,100) (0.9673307606,0) (1.073068652,100) (0.8991679018,100) (0.9317619509,0) (0.8792436823,100) (1.01234594,50) (1.027379312,50) (0.9253560652,50) (1.037211435,50) (0.8809439861,100) (1.081888464,100) (0.9639412159,100) (0.8784254358,50) (1.135069056,0) (1.114000679,100) (0.8296239648,100) (1.170411657,0) (1.087386042,0) (0.9888499547,100) (1.078032636,50) (0.8876284262,50)};
\addplot[forget plot,only marks,mark=diamond*,mark size=3.4pt,draw=foreignColor,fill=white,line width=0.8pt] coordinates {(1,70.96774194)};
\draw[draw=foreignColor,line width=1.2pt] (axis cs:0.79,100) -- (axis cs:1.21,100);
\nextgroupplot[xmin=0,xmax=1,ymin=0,ymax=1,xtick={0,0.2,0.4,0.6,0.8,1},ytick={0,0.2,0.4,0.6,0.8,1},xlabel={False-positive rate},ylabel={True-positive rate},]
\addplot[forget plot,black!45,dashed,line width=0.7pt] coordinates {(0,0) (1,1)};
\addplot[forget plot,rocColor,line width=1.2pt] coordinates {(0,0) (0.03225806452,0.1612903226) (0.2580645161,0.4193548387) (1,1)};
\node[anchor=south east,font=\small,fill=white,inner sep=3pt] at (axis cs:0.97,0.04) {AUC = 0.595};
\end{groupplot}
\end{tikzpicture}
\Description{A two-panel figure for planted error questions. The left panel shows own-paper scores as blue circles in the left column and unfamiliar-paper scores as orange squares in the right column on a 0 to 100 axis. A skipped or timed-out question is scored zero. Each point is one participant's arithmetic mean of the two questions in this category for that condition. Both observations from a participant use comparable small deterministic horizontal jitter and are not connected. White diamonds mark condition means and colored horizontal segments mark medians. The own condition has n=31, mean 85.5, and median 100.0; the unfamiliar condition has n=31, mean 71.0, and median 100.0. The right panel uses one minus the score proportion to predict the unfamiliar label, which is the positive class; its empirical ROC AUC is 0.595. The diagonal denotes chance-level discrimination. These are descriptive summaries only and do not support inferential or causal conclusions.}
\ifdefined\tikzexternalenable\tikzexternalenable\fi
\endgroup
    \caption{Distribution of the scores for the planted-error question family for participants' own papers and unfamiliar papers, and the ROC curve and AUC when classifying unfamiliar papers using only average scores on \textbf{planted error}.}
    \Description{A two-panel figure for planted-error questions. The left panel plots each participant's mean score for an own paper as a blue circle and for an unfamiliar paper as an orange square; scores occur at 0, 50, or 100 because each assessment contains two planted-error questions. Diamonds mark means of 85.5 for own papers and 71.0 for unfamiliar papers. The right panel shows the receiver-operating-characteristic curve for classifying unfamiliar papers from these scores. The area under the curve is 0.595, only slightly above the diagonal chance line.}
    \label{fig:planted_error_score_scatterplots}
\end{figure}
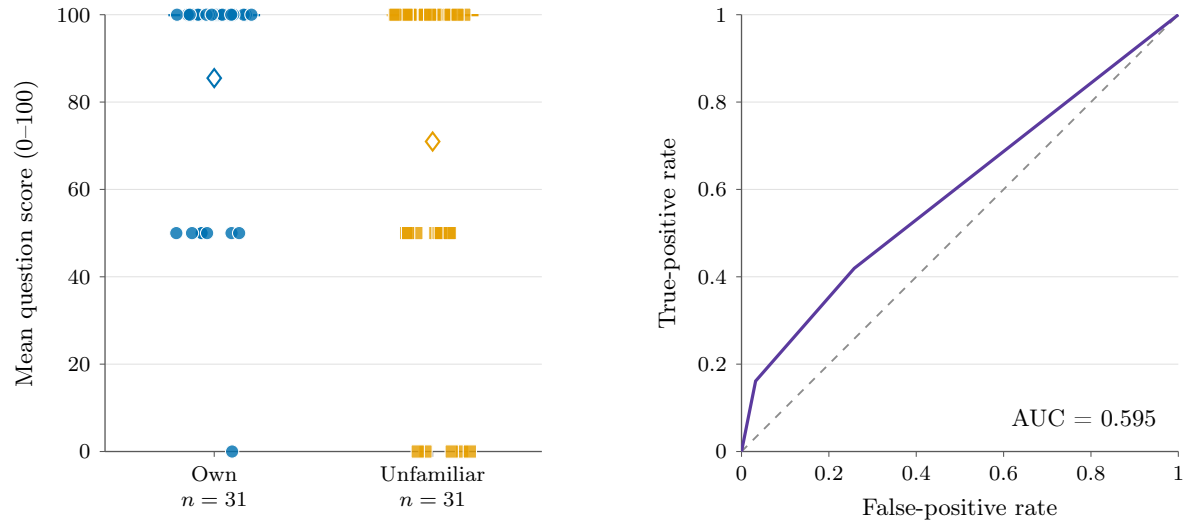

\begin{figure}[h]
    \centering
\begingroup
\ifdefined\tikzexternaldisable\tikzexternaldisable\fi
\definecolor{ownColor}{HTML}{0072B2}
\definecolor{foreignColor}{HTML}{E69F00}
\definecolor{rocColor}{HTML}{5B3A9E}
\begin{tikzpicture}
\begin{groupplot}[
group style={group size=2 by 1,horizontal sep=0.16\linewidth},
scale only axis,
width=0.35\linewidth,height=0.35\linewidth,
axis lines*=left,axis line style={black!65,line width=0.5pt},
tick align=outside,tick style={black!65,line width=0.4pt},
major tick length=2pt,scaled ticks=false,ymajorgrids=true,
grid style={black!12,line width=0.35pt},
label style={font=\small},tick label style={font=\footnotesize},
]
\nextgroupplot[xmin=-0.5,xmax=1.5,ymin=0,ymax=100,xtick={0,1},xticklabels={{Own\\$n=31$},{Unfamiliar\\$n=31$}},xticklabel style={align=center},ytick={0,20,40,60,80,100},ylabel={Mean question score (0--100)},]
\addplot[only marks,mark=*,mark size=2.5pt,draw=white,fill=ownColor,fill opacity=0.82] coordinates {(-0.1711902156,69) (-0.1157057426,39) (-0.1222714827,20) (0.1450974618,75) (-0.1035302449,49) (-0.1517703026,24.5) (0.1022644625,46) (-0.06673503759,42.5) (0.09045948558,29) (-0.06332137068,17) (0.1168025396,43.5) (-0.02427353041,40) (-0.003321029337,64) (0.1410472559,49) (-0.04602249743,60) (0.09648150676,19.5) (0.07534389074,15) (0.03198946403,28.5) (-0.0139487691,40) (0.01727408921,50.5) (0.1582909919,0) (0.1093793283,60.5) (0.107628179,35.5) (-0.04739450518,37.5) (0.07143246503,61.5) (0.04348200775,19) (-0.1783353174,10) (0.1342985732,34) (0.1236638816,21) (0.07289505231,18) (-0.1098821769,21.5)};
\addplot[forget plot,only marks,mark=diamond*,mark size=3.4pt,draw=ownColor,fill=white,line width=0.8pt] coordinates {(0,36.75806452)};
\draw[draw=ownColor,line width=1.2pt] (axis cs:-0.21,37.5) -- (axis cs:0.21,37.5);
\addplot[only marks,mark=square*,mark size=2.5pt,draw=white,fill=foreignColor,fill opacity=0.82] coordinates {(0.8288097844,25) (0.8842942574,4) (0.8777285173,0) (1.145097462,16.5) (0.8964697551,26.5) (0.8482296974,2) (1.102264462,6.5) (0.9332649624,11) (1.090459486,4) (0.9366786293,5) (1.11680254,26) (0.9757264696,9) (0.9966789707,19) (1.141047256,4) (0.9539775026,35) (1.096481507,19) (1.075343891,0) (1.031989464,30) (0.9860512309,8) (1.017274089,31.5) (1.158290992,4) (1.109379328,31.5) (1.107628179,0) (0.9526054948,22.5) (1.071432465,4) (1.043482008,0) (0.8216646826,20) (1.134298573,23) (1.123663882,4) (1.072895052,0) (0.8901178231,59)};
\addplot[forget plot,only marks,mark=diamond*,mark size=3.4pt,draw=foreignColor,fill=white,line width=0.8pt] coordinates {(1,14.51612903)};
\draw[draw=foreignColor,line width=1.2pt] (axis cs:0.79,9) -- (axis cs:1.21,9);
\nextgroupplot[xmin=0,xmax=1,ymin=0,ymax=1,xtick={0,0.2,0.4,0.6,0.8,1},ytick={0,0.2,0.4,0.6,0.8,1},xlabel={False-positive rate},ylabel={True-positive rate},]
\addplot[forget plot,black!45,dashed,line width=0.7pt] coordinates {(0,0) (1,1)};
\addplot[forget plot,rocColor,line width=1.2pt] coordinates {(0,0) (0.03225806452,0.1612903226) (0.03225806452,0.1935483871) (0.03225806452,0.3870967742) (0.03225806452,0.4193548387) (0.03225806452,0.4516129032) (0.03225806452,0.4838709677) (0.03225806452,0.5161290323) (0.06451612903,0.5161290323) (0.06451612903,0.5483870968) (0.09677419355,0.5483870968) (0.09677419355,0.5806451613) (0.1290322581,0.5806451613) (0.1612903226,0.5806451613) (0.1935483871,0.6451612903) (0.2258064516,0.6451612903) (0.2580645161,0.6774193548) (0.2903225806,0.6774193548) (0.3225806452,0.6774193548) (0.3225806452,0.7096774194) (0.3225806452,0.7419354839) (0.3548387097,0.7419354839) (0.3548387097,0.7741935484) (0.3548387097,0.8064516129) (0.3548387097,0.8387096774) (0.3870967742,0.8387096774) (0.4193548387,0.8387096774) (0.4193548387,0.8709677419) (0.4193548387,0.935483871) (0.4516129032,0.935483871) (0.4516129032,0.9677419355) (0.4838709677,0.9677419355) (0.5161290323,0.9677419355) (0.5483870968,0.9677419355) (0.6129032258,0.9677419355) (0.6451612903,0.9677419355) (0.6774193548,0.9677419355) (0.7096774194,0.9677419355) (0.7741935484,0.9677419355) (0.8064516129,0.9677419355) (0.8064516129,1) (0.8387096774,1) (0.8709677419,1) (0.9032258065,1) (0.935483871,1) (0.9677419355,1) (1,1)};
\node[anchor=south east,font=\small,fill=white,inner sep=3pt] at (axis cs:0.97,0.04) {AUC = 0.825};
\end{groupplot}
\end{tikzpicture}
\Description{A two-panel figure for unstated rationale questions. The left panel shows own-paper scores as blue circles in the left column and unfamiliar-paper scores as orange squares in the right column on a 0 to 100 axis. A skipped or timed-out question is scored zero. Each point is one participant's arithmetic mean of the two questions in this category for that condition. Both observations from a participant use comparable small deterministic horizontal jitter and are not connected. White diamonds mark condition means and colored horizontal segments mark medians. The own condition has n=31, mean 36.8, and median 37.5; the unfamiliar condition has n=31, mean 14.5, and median 9.0. The right panel uses one minus the score proportion to predict the unfamiliar label, which is the positive class; its empirical ROC AUC is 0.825. The diagonal denotes chance-level discrimination. These are descriptive summaries only and do not support inferential or causal conclusions.}
\ifdefined\tikzexternalenable\tikzexternalenable\fi
\endgroup
   \caption{Distribution of the scores for the \textbf{unstated-rationale} question family for participants' own papers and unfamiliar papers, and the ROC curve and AUC when using only average scores on unstated rationale.}
   \Description{A two-panel figure for unstated-rationale questions. The left panel plots participant mean scores from 0 to 100 for own papers as blue circles and unfamiliar papers as orange squares. Own-paper scores are generally higher, with means of 36.8 and 14.5, respectively; diamonds mark the means and horizontal lines mark medians. The right panel shows the receiver-operating-characteristic curve for classifying unfamiliar papers, with an area under the curve of 0.825 and a diagonal chance line.}
    \label{fig:unstated_rationale_score_scatterplots}
\end{figure}
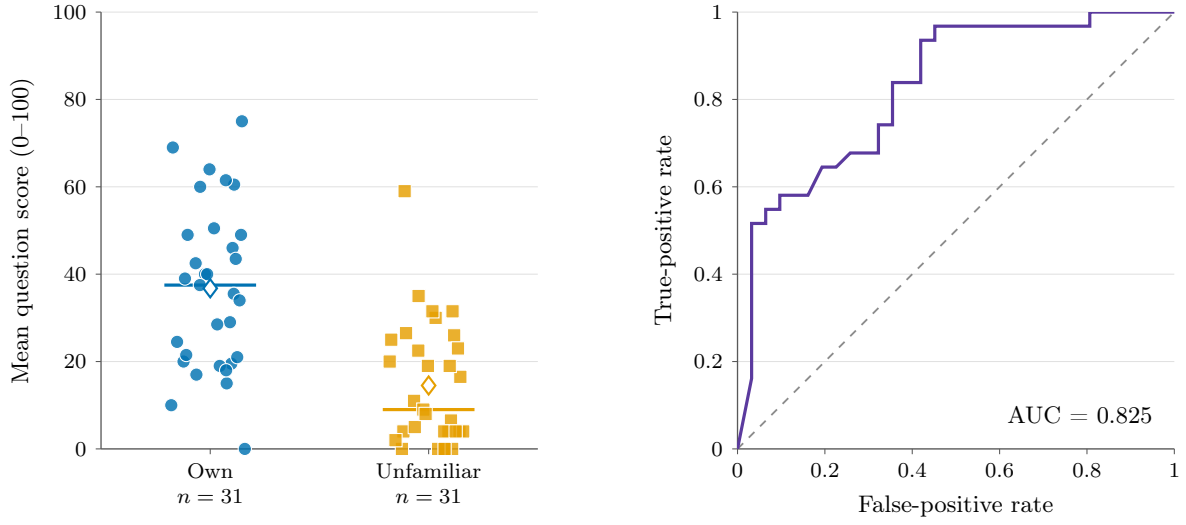

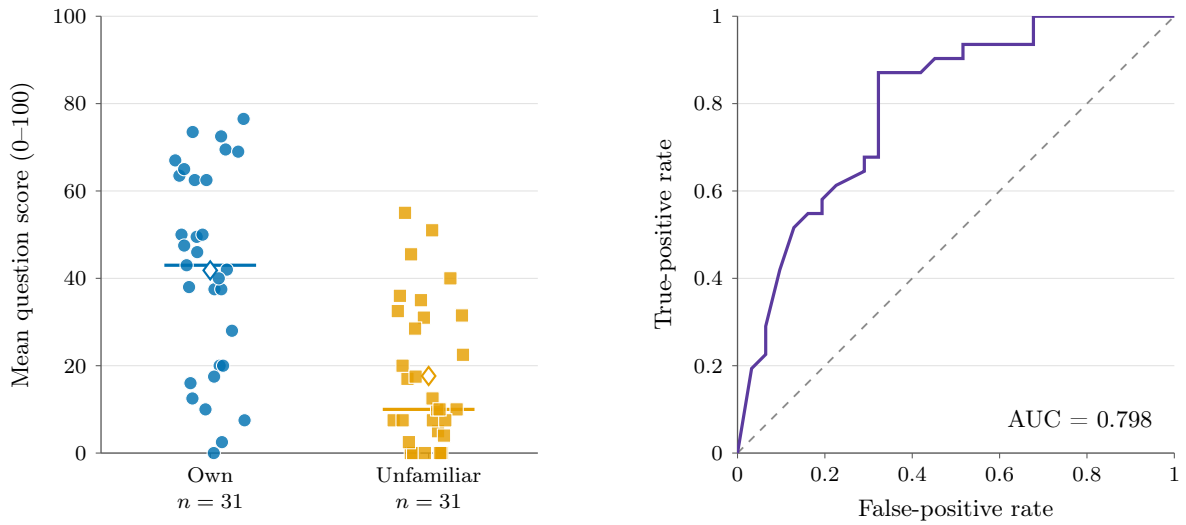
\begin{figure}[h]
    \centering
\begingroup
\ifdefined\tikzexternaldisable\tikzexternaldisable\fi
\definecolor{ownColor}{HTML}{0072B2}
\definecolor{foreignColor}{HTML}{E69F00}
\definecolor{rocColor}{HTML}{5B3A9E}
\begin{tikzpicture}
\begin{groupplot}[
group style={group size=2 by 1,horizontal sep=0.16\linewidth},
scale only axis,
width=0.35\linewidth,height=0.35\linewidth,
axis lines*=left,axis line style={black!65,line width=0.5pt},
tick align=outside,tick style={black!65,line width=0.4pt},
major tick length=2pt,scaled ticks=false,ymajorgrids=true,
grid style={black!12,line width=0.35pt},
label style={font=\small},tick label style={font=\footnotesize},
]
\nextgroupplot[xmin=-0.5,xmax=1.5,ymin=0,ymax=100,xtick={0,1},xticklabels={{Own\\$n=31$},{Unfamiliar\\$n=31$}},xticklabel style={align=center},ytick={0,20,40,60,80,100},ylabel={Mean question score (0--100)},]
\addplot[only marks,mark=*,mark size=2.5pt,draw=white,fill=ownColor,fill opacity=0.82] coordinates {(-0.02181580977,10) (0.04304544721,20) (0.1571642599,7.5) (-0.09693874327,38) (0.07024837114,69.5) (0.1281163201,69) (-0.08159152138,12.5) (0.1525793765,76.5) (0.01613421144,0) (-0.1315892342,50) (0.01940262537,37.5) (-0.0804209041,73.5) (-0.108170332,43) (0.05072058325,37.5) (-0.1601354353,67) (-0.07086119775,62.5) (-0.0621009078,49.5) (-0.03523595911,50) (0.05894338774,20) (0.09953655716,28) (0.05358250196,2.5) (-0.09044098793,16) (0.07619833929,42) (-0.1192006835,47.5) (0.03909061703,40) (-0.1411139569,63.5) (0.01776493038,17.5) (-0.01733553781,62.5) (-0.1193825027,65) (0.0499305353,72.5) (-0.05961504559,46)};
\addplot[forget plot,only marks,mark=diamond*,mark size=3.4pt,draw=ownColor,fill=white,line width=0.8pt] coordinates {(0,41.82258065)};
\draw[draw=ownColor,line width=1.2pt] (axis cs:-0.21,43) -- (axis cs:0.21,43);
\addplot[only marks,mark=square*,mark size=2.5pt,draw=white,fill=foreignColor,fill opacity=0.82] coordinates {(0.9781841902,31) (1.043045447,5) (1.15716426,22.5) (0.9030612567,17) (1.070248371,4) (1.12811632,10) (0.9184084786,0) (1.152579376,31.5) (1.016134211,51) (0.8684107658,36) (1.019402625,7.5) (0.9195790959,45.5) (0.891829668,55) (1.050720583,0) (0.8398645647,7.5) (0.9291388022,0) (0.9378990922,28.5) (0.9647640409,35) (1.058943388,0) (1.099536557,40) (1.053582502,0) (0.9095590121,2.5) (1.076198339,7.5) (0.8807993165,7.5) (1.039090617,10) (0.8588860431,32.5) (1.01776493,12.5) (0.9826644622,0) (0.8806174973,20) (1.049930535,10) (0.9403849544,17.5)};
\addplot[forget plot,only marks,mark=diamond*,mark size=3.4pt,draw=foreignColor,fill=white,line width=0.8pt] coordinates {(1,17.64516129)};
\draw[draw=foreignColor,line width=1.2pt] (axis cs:0.79,10) -- (axis cs:1.21,10);
\nextgroupplot[xmin=0,xmax=1,ymin=0,ymax=1,xtick={0,0.2,0.4,0.6,0.8,1},ytick={0,0.2,0.4,0.6,0.8,1},xlabel={False-positive rate},ylabel={True-positive rate},]
\addplot[forget plot,black!45,dashed,line width=0.7pt] coordinates {(0,0) (1,1)};
\addplot[forget plot,rocColor,line width=1.2pt] coordinates {(0,0) (0.03225806452,0.1935483871) (0.06451612903,0.2258064516) (0.06451612903,0.2580645161) (0.06451612903,0.2903225806) (0.09677419355,0.4193548387) (0.1290322581,0.5161290323) (0.1612903226,0.5483870968) (0.1935483871,0.5483870968) (0.1935483871,0.5806451613) (0.2258064516,0.6129032258) (0.2903225806,0.6451612903) (0.2903225806,0.6774193548) (0.3225806452,0.6774193548) (0.3225806452,0.7096774194) (0.3225806452,0.7419354839) (0.3225806452,0.7741935484) (0.3225806452,0.8064516129) (0.3225806452,0.8387096774) (0.3225806452,0.8709677419) (0.3870967742,0.8709677419) (0.4193548387,0.8709677419) (0.4516129032,0.9032258065) (0.4838709677,0.9032258065) (0.5161290323,0.9032258065) (0.5161290323,0.935483871) (0.5483870968,0.935483871) (0.5806451613,0.935483871) (0.6129032258,0.935483871) (0.6774193548,0.935483871) (0.6774193548,0.9677419355) (0.6774193548,1) (0.7419354839,1) (0.7741935484,1) (0.8064516129,1) (0.8387096774,1) (0.8709677419,1) (0.9032258065,1) (0.935483871,1) (0.9677419355,1) (1,1)};
\node[anchor=south east,font=\small,fill=white,inner sep=3pt] at (axis cs:0.97,0.04) {AUC = 0.798};
\end{groupplot}
\end{tikzpicture}
\Description{A two-panel figure for background knowledge questions. The left panel shows own-paper scores as blue circles in the left column and unfamiliar-paper scores as orange squares in the right column on a 0 to 100 axis. A skipped or timed-out question is scored zero. Each point is one participant's arithmetic mean of the two questions in this category for that condition. Both observations from a participant use comparable small deterministic horizontal jitter and are not connected. White diamonds mark condition means and colored horizontal segments mark medians. The own condition has n=31, mean 41.8, and median 43.0; the unfamiliar condition has n=31, mean 17.6, and median 10.0. The right panel uses one minus the score proportion to predict the unfamiliar label, which is the positive class; its empirical ROC AUC is 0.798. The diagonal denotes chance-level discrimination. These are descriptive summaries only and do not support inferential or causal conclusions.}
\ifdefined\tikzexternalenable\tikzexternalenable\fi
\endgroup
   \caption{Distribution of the scores for the background-knowledge question family for participants' own papers and unfamiliar papers, and the ROC curve and AUC when classifying unfamiliar papers using only average scores on \textbf{background knowledge}.}
    \Description{A two-panel figure for background-knowledge questions. The left panel plots participant mean scores from 0 to 100 for own papers as blue circles and unfamiliar papers as orange squares. Own-paper scores are generally higher, with means of 41.8 and 17.7, respectively; diamonds mark the means and horizontal lines mark medians. The right panel shows the receiver-operating-characteristic curve for classifying unfamiliar papers, with an area under the curve of 0.798 and a diagonal chance line.}
    \label{fig:background_knowledge_score_scatterplots}
\end{figure}

\begin{figure}[h]
    \centering
\begingroup
\ifdefined\tikzexternaldisable\tikzexternaldisable\fi
\definecolor{ownColor}{HTML}{0072B2}
\definecolor{foreignColor}{HTML}{E69F00}
\definecolor{rocColor}{HTML}{5B3A9E}
\begin{tikzpicture}
\begin{groupplot}[
group style={group size=2 by 1,horizontal sep=0.16\linewidth},
scale only axis,
width=0.35\linewidth,height=0.35\linewidth,
axis lines*=left,axis line style={black!65,line width=0.5pt},
tick align=outside,tick style={black!65,line width=0.4pt},
major tick length=2pt,scaled ticks=false,ymajorgrids=true,
grid style={black!12,line width=0.35pt},
label style={font=\small},tick label style={font=\footnotesize},
]
\nextgroupplot[xmin=-0.5,xmax=1.5,ymin=0,ymax=100,xtick={0,1},xticklabels={{Own\\$n=31$},{Unfamiliar\\$n=31$}},xticklabel style={align=center},ytick={0,20,40,60,80,100},ylabel={Mean question score (0--100)},]
\addplot[only marks,mark=*,mark size=2.5pt,draw=white,fill=ownColor,fill opacity=0.82] coordinates {(0.0721286625,0) (0.07567255896,7.5) (-0.174688735,14) (0.04827197984,39) (0.1634169878,2.5) (0.1587208864,27.5) (-0.1480271744,37.5) (-0.1013243126,0) (0.04786599352,73.5) (-0.007073808691,85) (-0.146258535,15) (-0.168449114,40) (0.1631411051,27.5) (0.03176501115,17.5) (-0.1390659162,47.5) (0.02683017037,50) (-0.028318608,81.5) (0.1582119098,27.5) (-0.1238288803,81) (0.07499084306,24) (-0.01433281441,39) (0.1788599209,55) (-0.1377418465,0) (-0.005373227708,36.5) (-0.1323072156,60) (-0.05134527332,57.5) (-0.007603902358,10) (-0.08316187704,54) (0.165017202,15) (-0.04012881454,0) (-0.1115427153,45)};
\addplot[forget plot,only marks,mark=diamond*,mark size=3.4pt,draw=ownColor,fill=white,line width=0.8pt] coordinates {(0,34.51612903)};
\draw[draw=ownColor,line width=1.2pt] (axis cs:-0.21,36.5) -- (axis cs:0.21,36.5);
\addplot[only marks,mark=square*,mark size=2.5pt,draw=white,fill=foreignColor,fill opacity=0.82] coordinates {(1.072128663,0) (1.075672559,0) (0.825311265,18.5) (1.04827198,8.5) (1.163416988,10) (1.158720886,0) (0.8519728256,0) (0.8986756874,0) (1.047865994,10) (0.9929261913,0) (0.853741465,0) (0.831550886,10) (1.163141105,0) (1.031765011,1) (0.8609340838,10) (1.02683017,0) (0.971681392,33.5) (1.15821191,0) (0.8761711197,0) (1.074990843,0) (0.9856671856,0) (1.178859921,0) (0.8622581535,0) (0.9946267723,0) (0.8676927844,0) (0.9486547267,0) (0.9923960976,0) (0.916838123,0) (1.165017202,0) (0.9598711855,0) (0.8884572847,76)};
\addplot[forget plot,only marks,mark=diamond*,mark size=3.4pt,draw=foreignColor,fill=white,line width=0.8pt] coordinates {(1,5.725806452)};
\draw[draw=foreignColor,line width=1.2pt] (axis cs:0.79,0) -- (axis cs:1.21,0);
\nextgroupplot[xmin=0,xmax=1,ymin=0,ymax=1,xtick={0,0.2,0.4,0.6,0.8,1},ytick={0,0.2,0.4,0.6,0.8,1},xlabel={False-positive rate},ylabel={True-positive rate},]
\addplot[forget plot,black!45,dashed,line width=0.7pt] coordinates {(0,0) (1,1)};
\addplot[forget plot,rocColor,line width=1.2pt] coordinates {(0,0) (0.1290322581,0.7096774194) (0.1290322581,0.7419354839) (0.1612903226,0.7419354839) (0.1935483871,0.7419354839) (0.1935483871,0.7741935484) (0.2258064516,0.9032258065) (0.2580645161,0.9032258065) (0.3225806452,0.9032258065) (0.3548387097,0.9032258065) (0.3548387097,0.935483871) (0.3870967742,0.935483871) (0.4838709677,0.935483871) (0.4838709677,0.9677419355) (0.5161290323,0.9677419355) (0.5483870968,0.9677419355) (0.6129032258,0.9677419355) (0.6451612903,0.9677419355) (0.6774193548,0.9677419355) (0.7096774194,0.9677419355) (0.7419354839,0.9677419355) (0.7741935484,0.9677419355) (0.8064516129,0.9677419355) (0.8387096774,0.9677419355) (0.8709677419,0.9677419355) (0.9032258065,0.9677419355) (0.9032258065,1) (0.935483871,1) (0.9677419355,1) (1,1)};
\node[anchor=south east,font=\small,fill=white,inner sep=3pt] at (axis cs:0.97,0.04) {AUC = 0.861};
\end{groupplot}
\end{tikzpicture}
\Description{A two-panel figure for failure mode questions. The left panel shows own-paper scores as blue circles in the left column and unfamiliar-paper scores as orange squares in the right column on a 0 to 100 axis. A skipped or timed-out question is scored zero. Each point is one participant's arithmetic mean of the two questions in this category for that condition. Both observations from a participant use comparable small deterministic horizontal jitter and are not connected. White diamonds mark condition means and colored horizontal segments mark medians. The own condition has n=31, mean 34.5, and median 36.5; the unfamiliar condition has n=31, mean 5.7, and median 0.0. The right panel uses one minus the score proportion to predict the unfamiliar label, which is the positive class; its empirical ROC AUC is 0.861. The diagonal denotes chance-level discrimination. These are descriptive summaries only and do not support inferential or causal conclusions.}
\ifdefined\tikzexternalenable\tikzexternalenable\fi
\endgroup
\caption{Distribution of the scores for the failure-mode question family for participants' own papers and unfamiliar papers, and the ROC curve and AUC when classifying unfamiliar papers using only average scores on \textbf{failure mode}.}
    \Description{A two-panel figure for failure-mode questions. The left panel plots participant mean scores from 0 to 100 for own papers as blue circles and unfamiliar papers as orange squares. Many unfamiliar-paper scores are zero, and own-paper scores are generally higher, with means of 34.5 and 5.7, respectively; diamonds mark the means and horizontal lines mark medians. The right panel shows the receiver-operating-characteristic curve for classifying unfamiliar papers, with an area under the curve of 0.861 and a diagonal chance line.}
    \label{fig:failure_mode_score_scatterplots}
\end{figure}
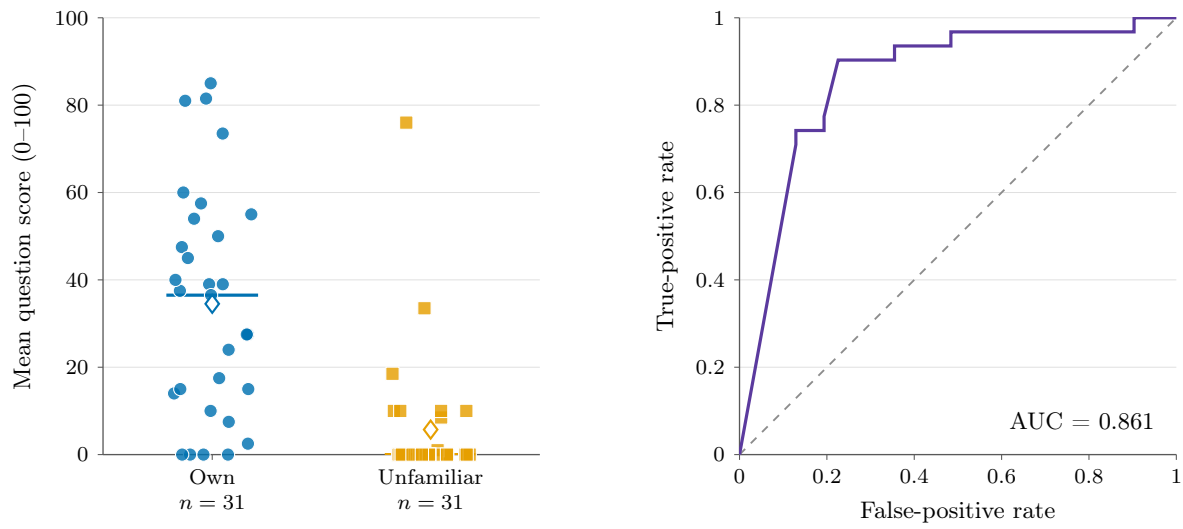

\clearpage
\section{Additional Statistical Tests}
\label{apx:full-stats}

\begin{figure}
    \centering
    \begingroup
\ifdefined\tikzexternaldisable\tikzexternaldisable\fi
\definecolor{ownColor}{HTML}{0072B2}
\definecolor{foreignColor}{HTML}{E69F00}

\newcommand{\scoreBox}[9]{%
  \path[
    draw=#8,
    fill=#8!12,
    pattern=#9,
    pattern color=#8,
    line width=0.7pt
  ]
    (axis cs:{#1-0.15},#2)
    rectangle
    (axis cs:{#1+0.15},#3);
  \draw[draw=#8,line width=0.7pt]
    (axis cs:#1,#5) -- (axis cs:#1,#2)
    (axis cs:#1,#3) -- (axis cs:#1,#6)
    (axis cs:{#1-0.084},#5) -- (axis cs:{#1+0.084},#5)
    (axis cs:{#1-0.084},#6) -- (axis cs:{#1+0.084},#6);
  \draw[black,line width=1pt]
    (axis cs:{#1-0.15},#4) -- (axis cs:{#1+0.15},#4);
  \addplot[
    forget plot,
    only marks,
    mark=diamond*,
    mark size=2.8pt,
    draw=#8,
    fill=white
  ] coordinates {(#1,#7)};
}

\newcommand{\scoreSignificance}[2]{%
  \draw[black,line width=0.6pt]
    (axis cs:{#1-0.18},103)
    -- (axis cs:{#1-0.18},106)
    -- (axis cs:{#1+0.18},106)
    -- (axis cs:{#1+0.18},103);
  \node[anchor=south,font=\normalsize,inner sep=1pt]
    at (axis cs:#1,106) {#2};
}

\begin{tikzpicture}
\begin{axis}[
  width=0.98\linewidth,
  height=0.48\linewidth,
  xmin=-0.48,xmax=3.48,
  ymin=0,ymax=113,
  ylabel={Score (0--100)},
  label style={font=\normalsize},
  tick label style={font=\normalsize},
  xtick={0,1,2,3},
  xticklabels={
    {Planted\\error},
    {Unstated\\rationale},
    {Background\\knowledge},
    {Failure\\mode}
  },
  xticklabel style={align=center},
  ytick={0,20,40,60,80,100},
  tick align=outside,
  axis lines*=left,
  ymajorgrids,
  grid style={draw=gray!25,line width=0.4pt},
  clip=false,
  legend style={
    at={(0.5,0)},
    anchor=north,
    yshift=-3.2em,
    draw=none,
    font=\normalsize,
    legend columns=3,
    /tikz/every even column/.append style={column sep=0.8em}
  },
]

\scoreBox{-0.18}{100}{100}{100}{100}{100}{85.483871}
  {ownColor}{north east lines}

\addplot[
  forget plot,only marks,mark=*,
  mark size=1.5pt,
  draw=ownColor,fill=ownColor,fill opacity=0.5
] coordinates {(-0.18,0) (-0.18,0) (-0.18,0) (-0.18,0) (-0.18,0) (-0.18,0) (-0.18,0) (-0.18,0) (-0.18,0)};
\scoreBox{0.18}{0}{100}{100}{0}{100}{70.967742}
  {foreignColor}{north west lines}
\scoreSignificance{0}{*}

\scoreBox{0.82}{16.25}{50}{37}{0}{90}{36.758065}
  {ownColor}{north east lines}
\scoreBox{1.18}{0}{27.25}{8}{0}{62}{14.516129}
  {foreignColor}{north west lines}
\scoreSignificance{1}{***}

\scoreBox{1.82}{13.5}{65.75}{44}{0}{100}{41.822581}
  {ownColor}{north east lines}
\scoreBox{2.18}{0}{34}{9}{0}{70}{17.645161}
  {foreignColor}{north west lines}
\scoreSignificance{2}{***}

\scoreBox{2.82}{0}{69.5}{22.5}{0}{100}{34.516129}
  {ownColor}{north east lines}
\scoreBox{3.18}{0}{0}{0}{0}{0}{5.7258065}
  {foreignColor}{north west lines}

\addplot[
  forget plot,only marks,mark=*,
  mark size=1.5pt,
  draw=foreignColor,fill=foreignColor,fill opacity=0.5
] coordinates {(3.18,2) (3.18,62) (3.18,90) (3.18,15) (3.18,5) (3.18,20) (3.18,37) (3.18,67) (3.18,20) (3.18,20) (3.18,5) (3.18,12)};
\scoreSignificance{3}{***}

\addlegendimage{
  legend image code/.code={
    \draw[
      draw=ownColor,
      fill=ownColor!12,
      pattern=north east lines,
      pattern color=ownColor
    ] (0cm,-0.09cm) rectangle (0.34cm,0.09cm);
  }
}
\addlegendentry{Own paper ($n=62$)}

\addlegendimage{
  legend image code/.code={
    \draw[
      draw=foreignColor,
      fill=foreignColor!12,
      pattern=north west lines,
      pattern color=foreignColor
    ] (0cm,-0.09cm) rectangle (0.34cm,0.09cm);
  }
}
\addlegendentry{Unfamiliar paper ($n=62$)}

\addlegendimage{
  only marks,mark=diamond*,mark size=2.8pt,
  draw=black,fill=white
}
\addlegendentry{Mean}

\end{axis}
\end{tikzpicture}
\ifdefined\tikzexternalenable\tikzexternalenable\fi
\endgroup
    \caption{Score distributions by question family and paper type ($n=62$ per box). Diamonds indicate means; brackets mark significant differences using Holm-adjusted $p$-values (* $p<.05$, *** $p<.001$).}
    \Description{Four pairs of box plots which compare scores from 0 to 100 for own and unfamiliar papers across planted error, unstated rationale, background knowledge, and failure mode questions. Own-paper scores are higher on average in every category, with means of 85.5 versus 71.0, 36.8 versus 14.5, 41.8 versus 17.7, and 34.5 versus 5.7, respectively. All four comparisons are statistically significant after Holm adjustment. Colors and hatch patterns distinguish paper types, and diamonds mark means.}
    \label{fig:h1_effects}
\end{figure}
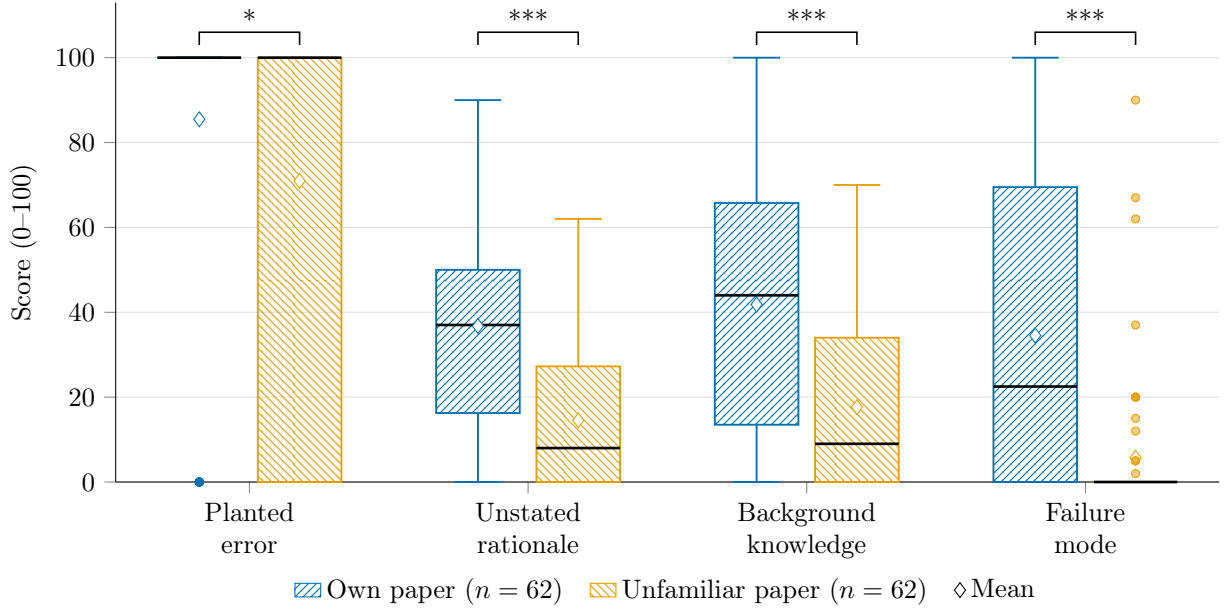

For further analysis we define $\mathrm{H}_1$ to be the null hypothesis that, within each question family, participants cannot significantly be separated based on whether they were assessed on their own paper vs. on an unfamiliar paper.
To test this hypothesis, we use the model: \texttt{score $\sim$ paper\_type * question\_family + (1 | participant\_id)}.
We fit it on all $496$ question responses, from $2$ tests with $8$ questions each over $31$ participants. Questions which were unanswered due to skipping or test time-out were marked as having a score of $0$. Details of this data are shown in \cref{tab:question-score-descriptives}.

We normalize scores to the range $[0, 1]$ by dividing by $100$. Scores are bounded and clustered around $0$ and $1$, especially for the planted-error question family. We therefore fit an ordered-beta regression model to our data \cite{kubinec2023ordered}. For each question family $q$, the maximum likelihood ordered-beta regression model fits logit-link values $\beta_{\text{own}, q}$ and $\beta_{\text{unfamiliar}, q}$ indicating the connections between the question family and score for each paper type.

For each question family $q$, we construct the statistic $C_q \doteq \beta_{\text{own},q} - \beta_{\text{unfamiliar},q}$. Our null hypothesis for each question family $q$ is $H_0 : C_q = 0$. We calculate marginal means using the ``emmeans'' R package \cite{searle1980population}, and correct for multiple comparisons using $4$-way Holm correction. \Cref{tab:h1-own-foreign-contrasts,tab:h1-estimated-marginal-means,fig:h1_effects} show the full statistical results of the test.

Our statistical analysis rejects the null hypothesis for all four question families under Holm correction.
We find that $p$-values are significant at the $0.05$ level for planted error, and at the $0.001$ level for all free-response question families. 
This is strong evidence to indicate that the four question families are predictive of the capacity to verify one's paper.

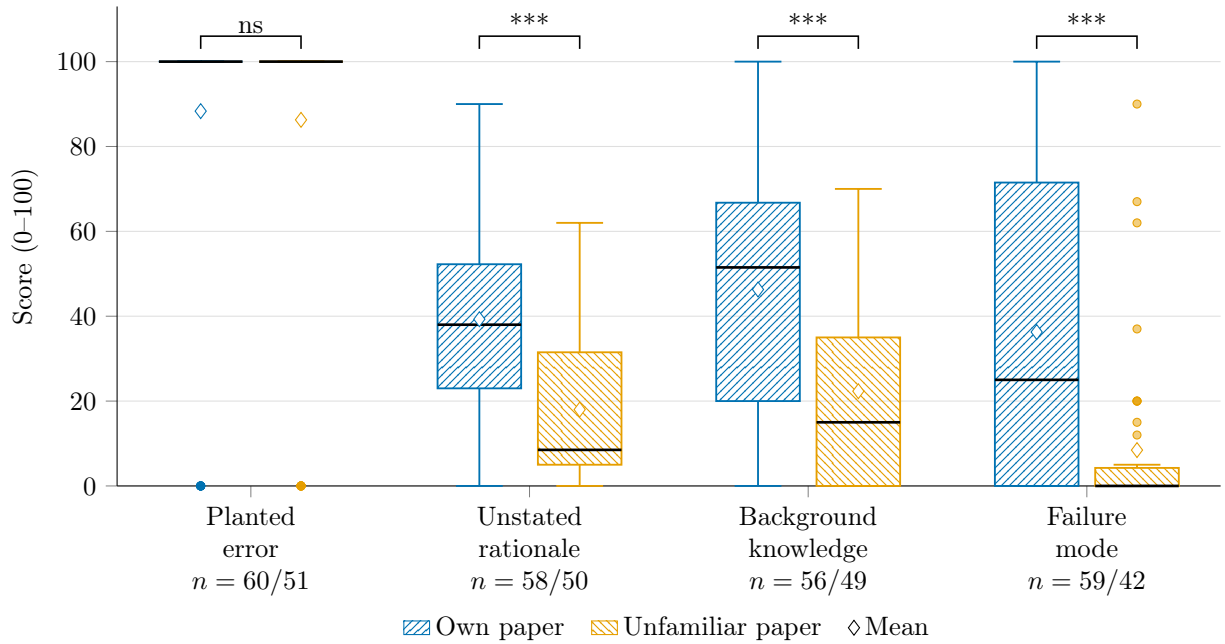
\begin{figure}[t]
\centering
\begingroup
\ifdefined\tikzexternaldisable\tikzexternaldisable\fi
\definecolor{ownColor}{HTML}{0072B2}
\definecolor{foreignColor}{HTML}{E69F00}

\newcommand{\answeredBox}[9]{%
  \path[
    draw=#8,
    fill=#8!12,
    pattern=#9,
    pattern color=#8,
    line width=0.7pt
  ]
    (axis cs:{#1-0.15},#2)
    rectangle
    (axis cs:{#1+0.15},#3);
  \draw[draw=#8,line width=0.7pt]
    (axis cs:#1,#5) -- (axis cs:#1,#2)
    (axis cs:#1,#3) -- (axis cs:#1,#6)
    (axis cs:{#1-0.084},#5) -- (axis cs:{#1+0.084},#5)
    (axis cs:{#1-0.084},#6) -- (axis cs:{#1+0.084},#6);
  \draw[black,line width=1pt]
    (axis cs:{#1-0.15},#4) -- (axis cs:{#1+0.15},#4);
  \addplot[
    forget plot,
    only marks,
    mark=diamond*,
    mark size=2.8pt,
    draw=#8,
    fill=white
  ] coordinates {(#1,#7)};
}

\newcommand{\answeredSignificance}[2]{%
  \draw[black,line width=0.6pt]
    (axis cs:{#1-0.18},103)
    -- (axis cs:{#1-0.18},106)
    -- (axis cs:{#1+0.18},106)
    -- (axis cs:{#1+0.18},103);
  \node[anchor=south,font=\normalsize,inner sep=1pt]
    at (axis cs:#1,106) {#2};
}

\begin{tikzpicture}
\begin{axis}[
  width=0.98\linewidth,
  height=0.48\linewidth,
  xmin=-0.48,xmax=3.48,
  ymin=0,ymax=113,
  ylabel={Score (0--100)},
  label style={font=\normalsize},
  tick label style={font=\normalsize},
  xtick={0,1,2,3},
  xticklabels={
    {Planted\\error\\$n=60/51$},
    {Unstated\\rationale\\$n=58/50$},
    {Background\\knowledge\\$n=56/49$},
    {Failure\\mode\\$n=59/42$}
  },
  xticklabel style={align=center},
  ytick={0,20,40,60,80,100},
  tick align=outside,
  axis lines*=left,
  ymajorgrids,
  grid style={draw=gray!25,line width=0.4pt},
  clip=false,
  legend style={
    at={(0.5,0)},
    anchor=north,
    yshift=-4.5em,
    draw=none,
    font=\normalsize,
    legend columns=3,
    /tikz/every even column/.append style={column sep=0.8em}
  },
]

\answeredBox{-0.18}{100}{100}{100}{100}{100}{88.333333}
  {ownColor}{north east lines}

\addplot[
  forget plot,only marks,mark=*,
  mark size=1.5pt,
  draw=ownColor,fill=ownColor,fill opacity=0.5
] coordinates {(-0.18,0) (-0.18,0) (-0.18,0) (-0.18,0) (-0.18,0) (-0.18,0) (-0.18,0)};
\answeredBox{0.18}{100}{100}{100}{100}{100}{86.27451}
  {foreignColor}{north west lines}

\addplot[
  forget plot,only marks,mark=*,
  mark size=1.5pt,
  draw=foreignColor,fill=foreignColor,fill opacity=0.5
] coordinates {(0.18,0) (0.18,0) (0.18,0) (0.18,0) (0.18,0) (0.18,0) (0.18,0)};
\answeredSignificance{0}{ns}

\answeredBox{0.82}{23}{52.25}{38}{0}{90}{39.293103}
  {ownColor}{north east lines}
\answeredBox{1.18}{5}{31.5}{8.5}{0}{62}{18}
  {foreignColor}{north west lines}
\answeredSignificance{1}{***}

\answeredBox{1.82}{20}{66.75}{51.5}{0}{100}{46.303571}
  {ownColor}{north east lines}
\answeredBox{2.18}{0}{35}{15}{0}{70}{22.326531}
  {foreignColor}{north west lines}
\answeredSignificance{2}{***}

\answeredBox{2.82}{0}{71.5}{25}{0}{100}{36.271186}
  {ownColor}{north east lines}
\answeredBox{3.18}{0}{4.25}{0}{0}{5}{8.452381}
  {foreignColor}{north west lines}

\addplot[
  forget plot,only marks,mark=*,
  mark size=1.5pt,
  draw=foreignColor,fill=foreignColor,fill opacity=0.5
] coordinates {(3.18,62) (3.18,90) (3.18,15) (3.18,20) (3.18,37) (3.18,67) (3.18,20) (3.18,20) (3.18,12)};
\answeredSignificance{3}{***}

\addlegendimage{
  legend image code/.code={
    \draw[
      draw=ownColor,
      fill=ownColor!12,
      pattern=north east lines,
      pattern color=ownColor
    ] (0cm,-0.09cm) rectangle (0.34cm,0.09cm);
  }
}
\addlegendentry{Own paper}

\addlegendimage{
  legend image code/.code={
    \draw[
      draw=foreignColor,
      fill=foreignColor!12,
      pattern=north west lines,
      pattern color=foreignColor
    ] (0cm,-0.09cm) rectangle (0.34cm,0.09cm);
  }
}
\addlegendentry{Unfamiliar paper}

\addlegendimage{
  only marks,mark=diamond*,mark size=2.8pt,
  draw=black,fill=white
}
\addlegendentry{Mean}

\end{axis}
\end{tikzpicture}
\ifdefined\tikzexternalenable\tikzexternalenable\fi
\endgroup

\caption{Score distributions for answered questions only, grouped by question family and paper type. Sample sizes show the counts of responses to question sets based on own/unfamiliar paper. Diamonds indicate means. Brackets report Holm-adjusted comparisons (*** $p<.001$; ns, not significant).}
\label{fig:h1_answered_only_score_distribution}

\Description{Four pairs of box plots compare answered-question scores from 0 to 100 for own and unfamiliar papers. Sample sizes for own and unfamiliar responses are 60 and 51 for planted error, 58 and 50 for unstated rationale, 56 and 49 for background knowledge, and 59 and 42 for failure mode. The planted-error distributions are similarly concentrated near 100 and do not differ significantly. Own-paper scores are higher for each free-response family, and all three differences are significant at the 0.001 level after Holm adjustment. Blue and orange hatch patterns distinguish paper types, and diamonds mark means.}

\end{figure}

We also create an alternative model in which we exclude all skipped and timed-out questions from our analyzed data.
The results for this are reported in~\cref{tab:sensitivity_answered_h1_contrasts,tab:h1-answered-only-marginal-means,fig:h1_answered_only_score_distribution}.
Under this model, all free-response question families show significant contrasts between scores in the own and the unfamiliar settings ($p < 0.001$; Holm-adjusted).
Nevertheless, the effect fails to meet the threshold for significance on planted error (multiple choice) questions. 
For this question family, this additional result suggests a different driver behind the previously witnessed effect: participants seem more likely to skip or not reach the planted error questions in the unfamiliar-paper setting.
When they do answer these questions, they are not significantly less accurate than in the own-paper setting.

We also present additional results for $\mathrm{H}_2$, the null hypothesis that there are no significant differences between in-field and out-of-field unfamiliar paper scores. These results are described in~\cref{tab:h2-field-contrast,tab:h2-foreign-paper-marginal-means,tab:h2-participant-foreign-field-robustness}.

\begin{table}[h]
\centering
\caption{Descriptive metrics of question-scores overall, grouped by paper ownership and by question family. Scores range from 0 to 100. For every group, the total number of observations equals \(n\), and no scores are missing. Zero, Full, and Nonresponse denote the percentages of observations with a score of 0, a score of 100, and no response, respectively.}
\Description{A descriptive-statistics table reporting sample size, mean, standard deviation, median, and percentages of zero, full-credit, and unanswered scores. Results are shown overall and by paper ownership and question family. Across 496 responses, the overall mean is 38.43. Own-paper responses average 49.65 versus 27.21 for unfamiliar-paper responses. Planted-error questions have the highest mean, 78.23, while failure-mode questions have the lowest, 20.12. Zero and nonresponse rates are consistently higher for unfamiliar papers.}
\label{tab:question-score-descriptives}
\small
\setlength{\tabcolsep}{4pt}
\renewcommand{\arraystretch}{1.1}
\begin{tabular*}{\linewidth}{
  @{\extracolsep{\fill}}lrrrrrrr@{}
}
\toprule
& & \multicolumn{3}{c}{Score}
  & \multicolumn{3}{c}{Rate (\%)} \\
\cmidrule(lr){3-5}\cmidrule(l){6-8}
Group & \(n\) & Mean & SD & Median
      & Zero & Full & Nonresponse \\
\midrule
Overall & 496 & 38.43 & 39.85 & 25.0 & 34.1 & 21.0 & 14.3 \\

\addlinespace
\multicolumn{8}{@{}l}{\textit{By paper ownership}} \\
\quad Unfamiliar paper & 248 & 27.21 & 38.05 & 5.0 & 46.8 & 17.7 & 22.6 \\
\quad Own paper & 248 & 49.65 & 38.50 & 50.0 & 21.4 & 24.2 & 6.0 \\

\addlinespace
\multicolumn{8}{@{}l}{\textit{By question family}} \\
\quad Planted error & 124 & 78.23 & 41.44 & 100.0 & 21.8 & 78.2 & 10.5 \\
\quad Unstated rationale & 124 & 25.64 & 24.36 & 20.0 & 21.0 & 0.0 & 12.9 \\
\quad Background knowledge & 124 & 29.73 & 29.69 & 20.0 & 33.1 & 2.4 & 15.3 \\
\quad Failure mode & 124 & 20.12 & 31.97 & 0.0 & 60.5 & 3.2 & 18.5 \\

\addlinespace
\multicolumn{8}{@{}l}{\textit{Unfamiliar paper, by question family}} \\
\quad Planted error & 62 & 70.97 & 45.76 & 100.0 & 29.0 & 71.0 & 17.7 \\
\quad Unstated rationale & 62 & 14.52 & 17.76 & 8.0 & 32.3 & 0.0 & 19.4 \\
\quad Background knowledge & 62 & 17.65 & 22.23 & 9.0 & 45.2 & 0.0 & 21.0 \\
\quad Failure mode & 62 & 5.73 & 16.89 & 0.0 & 80.6 & 0.0 & 32.3 \\

\addlinespace
\multicolumn{8}{@{}l}{\textit{Own paper, by question family}} \\
\quad Planted error & 62 & 85.48 & 35.51 & 100.0 & 14.5 & 85.5 & 3.2 \\
\quad Unstated rationale & 62 & 36.76 & 25.10 & 37.0 & 9.7 & 0.0 & 6.5 \\
\quad Background knowledge & 62 & 41.82 & 31.39 & 44.0 & 21.0 & 4.8 & 9.7 \\
\quad Failure mode & 62 & 34.52 & 36.80 & 22.5 & 40.3 & 6.5 & 4.8 \\
\bottomrule
\end{tabular*}
\end{table}



\begin{table}[h]
\centering
\caption{Estimated marginal mean scores for $\mathrm{H}_1$ by question family and paper ownership. Scores are reported on a 0--100 scale, with 95\% confidence intervals (CIs).}
\Description{A table of model-estimated marginal mean scores and 95-percent confidence intervals for each question family, split by unfamiliar and own papers. Estimated means are higher for own papers in all four families: 97.19 versus 91.33 for planted error, 42.79 versus 26.81 for unstated rationale, 49.30 versus 31.36 for background knowledge, and 47.01 versus 14.43 for failure mode.}
\label{tab:h1-estimated-marginal-means}
\small
\setlength{\tabcolsep}{4pt}
\renewcommand{\arraystretch}{1.1}
\begin{tabular*}{\linewidth}{
  @{\extracolsep{\fill}}lrlrl@{}
}
\toprule
& \multicolumn{2}{c}{Unfamiliar paper}
& \multicolumn{2}{c}{Own paper} \\
\cmidrule(lr){2-3}\cmidrule(l){4-5}
Question family & Mean & 95\% CI & Mean & 95\% CI \\
\midrule
Planted error
  & 91.33
  & [83.99, 95.49]
  & 97.19
  & [93.89, 98.74] \\
Unstated rationale
  & 26.81
  & [21.82, 32.46]
  & 42.79
  & [37.13, 48.64] \\
Background knowledge
  & 31.36
  & [25.44, 37.96]
  & 49.30
  & [43.03, 55.60] \\
Failure mode
  & 14.43
  & [9.46, 21.39]
  & 47.01
  & [39.95, 54.19] \\
\bottomrule
\end{tabular*}
\end{table}
\begin{table}[h]
\centering
\caption{Contrasts calculated for $\mathrm{H}_1$ between own vs. unfamiliar paper conditions on the model's link scale. Positive estimates indicate higher scores for own papers. We apply Holm adjustment across the four contrasts. Bold estimates and stars indicate significance using the Holm-adjusted \(p\)-values: \({}^{***}p<.001\), \({}^{**}p<.01\), \({}^{*}p<.05\).}
\Description{A table of own-minus-unfamiliar contrasts on the ordered-beta model's link scale for the four question families. Columns report the estimate, standard error, raw p-value, and Holm-adjusted p-value. All estimates are positive and statistically significant after adjustment: 1.190 for planted error, 0.714 for unstated rationale, 0.755 for background knowledge, and 1.660 for failure mode.}
\label{tab:h1-own-foreign-contrasts}
\small
\setlength{\tabcolsep}{4pt}
\renewcommand{\arraystretch}{1.1}
\begin{tabular*}{\linewidth}{
  @{\extracolsep{\fill}}lrrrr@{}
}
\toprule
& & & \multicolumn{2}{c}{\(p\)-value} \\
\cmidrule(l){4-5}
Question family & Estimate & SE & Raw & Holm-adjusted \\
\midrule
Planted error
  & \(\mathbf{1.190}^{*}\)
  & 0.475
  & 0.0122
  & 0.0122 \\
Unstated rationale
  & \(\mathbf{0.714}^{***}\)
  & 0.165
  & \(<0.0001\)
  & \(<0.0001\) \\
Background knowledge
  & \(\mathbf{0.755}^{***}\)
  & 0.181
  & \(<0.0001\)
  & \(<0.0001\) \\
Failure mode
  & \(\mathbf{1.660}^{***}\)
  & 0.267
  & \(<0.0001\)
  & \(<0.0001\) \\
\bottomrule
\end{tabular*}
\end{table}

\begin{table}[h]
\centering
\caption{Estimated marginal mean scores for $\mathrm{H}_1$ using questions that were answered only, grouped by question family and paper ownership. Scores are reported on a 0--100 scale, with 95\% confidence intervals (CIs).}
\Description{A table of model-estimated marginal mean scores and 95-percent confidence intervals using answered questions only, split by question family and paper ownership. Planted-error means are nearly identical for unfamiliar and own papers, 98.08 and 98.44. Own-paper means remain higher for the three free-response families: 42.34 versus 27.24 for unstated rationale, 49.80 versus 32.45 for background knowledge, and 46.18 versus 15.79 for failure mode.}
\label{tab:h1-answered-only-marginal-means}
\small
\setlength{\tabcolsep}{4pt}
\renewcommand{\arraystretch}{1.1}
\begin{tabular*}{\linewidth}{
  @{\extracolsep{\fill}}lrlrl@{}
}
\toprule
& \multicolumn{2}{c}{Unfamiliar paper}
& \multicolumn{2}{c}{Own paper} \\
\cmidrule(lr){2-3}\cmidrule(l){4-5}
Question family & Mean & 95\% CI & Mean & 95\% CI \\
\midrule
Planted error
  & 98.08
  & [95.26, 99.24]
  & 98.44
  & [96.17, 99.38] \\
Unstated rationale
  & 27.24
  & [22.13, 33.04]
  & 42.34
  & [36.62, 48.28] \\
Background knowledge
  & 32.45
  & [26.27, 39.31]
  & 49.80
  & [43.38, 56.22] \\
Failure mode
  & 15.79
  & [10.17, 23.70]
  & 46.18
  & [38.98, 53.55] \\
\bottomrule
\end{tabular*}
\end{table}



\begin{table}[h]
\centering
\caption{Contrasts calculated for $\mathrm{H}_1$ between own vs. unfamiliar paper conditions on the model's link scale but using answered questions only. Skipped, timed-out, and unknown-status questions are excluded. Positive estimates indicate higher scores for own papers. Holm adjustment is applied across the four contrasts. Bold estimates and stars indicate significance using Holm-adjusted \(p\)-values: \({}^{***}p<.001\), \({}^{**}p<.01\), \({}^{*}p<.05\).}
\Description{A table of own-minus-unfamiliar contrasts on the model's link scale after excluding skipped, timed-out, and unknown-status questions. Columns report the estimate, standard error, raw p-value, and Holm-adjusted p-value. The planted-error contrast is small and not significant, with estimate 0.210 and adjusted p-value 0.7171. The unstated-rationale, background-knowledge, and failure-mode contrasts are positive and significant after Holm adjustment, with estimates 0.674, 0.725, and 1.521.}
\label{tab:sensitivity_answered_h1_contrasts}
\small
\setlength{\tabcolsep}{4pt}
\renewcommand{\arraystretch}{1.1}
\begin{tabular*}{\linewidth}{
  @{\extracolsep{\fill}}lrrrr@{}
}
\toprule
& & & \multicolumn{2}{c}{\(p\)-value} \\
\cmidrule(l){4-5}
Question family & Estimate & SE & Raw & Holm-adjusted \\
\midrule
Planted error
  & \(0.210\)
  & 0.580
  & 0.7171
  & 0.7171 \\
Unstated rationale
  & \(\mathbf{0.674}^{***}\)
  & 0.170
  & \(<0.0001\)
  & 0.0002 \\
Background knowledge
  & \(\mathbf{0.725}^{***}\)
  & 0.189
  & 0.0001
  & 0.0002 \\
Failure mode
  & \(\mathbf{1.521}^{***}\)
  & 0.283
  & \(<0.0001\)
  & \(<0.0001\) \\
\bottomrule
\end{tabular*}
\end{table}



\begin{table}[h]
\centering
\caption{Estimated marginal mean scores for $\mathrm{H}_2$ on unfamiliar papers by whether their field matches that of the participant. Scores are reported on a 0--100 scale, with 95\% confidence intervals (CIs).}
\Description{A two-row table comparing estimated marginal mean scores on unfamiliar papers by field match. The out-of-field mean is 36.77 with a 95-percent confidence interval from 28.22 to 46.23. The in-field mean is 40.55 with a confidence interval from 31.93 to 49.79. The intervals overlap substantially.}
\label{tab:h2-foreign-paper-marginal-means}
\small
\setlength{\tabcolsep}{4pt}
\renewcommand{\arraystretch}{1.1}
\begin{tabular*}{\linewidth}{
  @{\extracolsep{\fill}}lrl@{}
}
\toprule
Field match & Mean & 95\% CI \\
\midrule
Out-of-field & 36.77 & [28.22, 46.23] \\
In-field & 40.55 & [31.93, 49.79] \\
\bottomrule
\end{tabular*}
\end{table}




\begin{table}[h]
\centering
\caption{$\mathrm{H}_2$ in-field vs. out-field contrast on the model's link scale for unfamiliar papers. A positive estimate indicates higher scores for in-field papers.}
\Description{A one-row table reporting the in-field-minus-out-of-field contrast for unfamiliar-paper scores on the model's link scale. The estimate is 0.160 with standard error 0.205 and p-value 0.4375, providing no statistically significant evidence of a field-match effect.}
\label{tab:h2-field-contrast}
\small
\setlength{\tabcolsep}{4pt}
\renewcommand{\arraystretch}{1.1}
\begin{tabular*}{\linewidth}{
  @{\extracolsep{\fill}}rrr@{}
}
\toprule
Estimate & SE & \(p\)-value \\
\midrule
0.160 & 0.205 & 0.4375 \\
\bottomrule
\end{tabular*}
\end{table}


\begin{table}[h]
\centering
\caption{Participant-level robustness tests comparing unfamiliar-paper scores by whether they match the field of the participant. Confidence intervals (CIs) are reported at the 95\% level.}
\Description{A two-row table reporting participant-level robustness tests for differences between in-field and out-of-field unfamiliar-paper scores. The Welch two-sample t-test has statistic negative 0.67, p-value 0.5084, and a 95-percent confidence interval from negative 12.808 to 6.491. The Wilcoxon rank-sum test has statistic 100, p-value 0.4407, and confidence interval from negative 12.750 to 6.125. Neither test finds a statistically significant difference.}
\label{tab:h2-participant-foreign-field-robustness}
\small
\setlength{\tabcolsep}{4pt}
\renewcommand{\arraystretch}{1.1}
\begin{tabular*}{\linewidth}{
  @{\extracolsep{\fill}}lrrl@{}
}
\toprule
Test & Statistic & \(p\)-value & 95\% CI \\
\midrule
Welch two-sample \(t\)-test
  &-0.67 & 0.5084 & [$-12.808$, $6.491$] \\
Wilcoxon rank-sum test
  & 100.00 & 0.4407 & [$-12.750$, $6.125$] \\
\bottomrule
\end{tabular*}
\end{table}


\clearpage
\section{Deductive Codebook}
\label{apx:deductive_codebook}

\Cref{tab:deductive-codebook} shows our deductive codebook with occurrence counts after coding. 

\begin{table*}[h]
\centering
\caption{Our final deductive codebook with columns for the \textbf{Code} name, its \textbf{Description}, the different \textbf{Type and values} into which we have subdivided each code, and the total number of occurrence \textbf{Count} across all transcripts.}
\Description{A four-column table containing 18 deductive interview codes. For each code, the table provides its name, operational description, categorical or binary values, and total occurrence count across transcripts. The most frequent codes are perceived validity with 91 occurrences, question clarity with 79, question relevance with 71, background familiarity with 70, and question difficulty with 64. The least frequent are grading false positive with 7 occurrences and grading false negative with 6.}
\label{tab:deductive-codebook}
\small
\setlength{\tabcolsep}{3pt}
\renewcommand{\arraystretch}{1.08}
\newcommand{\cbvalues}[2]{\textit{#1:} #2}
\begin{tabularx}{\textwidth}{@{}
  >{\RaggedRight\arraybackslash}p{0.24\textwidth}
  >{\RaggedRight\arraybackslash}X
  >{\RaggedRight\arraybackslash}p{0.27\textwidth}r@{}}
\toprule
\textbf{Code} & \textbf{Description} & \textbf{Type and values} & \textbf{Count} \\
\midrule
Perceived validity
  & The participant perceived the system as providing valid evidence to the fact that someone has understood the paper.
  & \cbvalues{Category}{Disagree, Neither agree nor disagree, Agree} & 91 \\
\addlinespace[2pt]
Question clarity
  & Whether the wording and requested response were understandable.
  & \cbvalues{Category}{Unclear; Partly clear; Clear} & 79 \\
\addlinespace[2pt]
Question relevance
  & The participant found the question relevant.
  & \cbvalues{Category}{Disagree, Neither agree nor disagree, Agree} & 71 \\
\addlinespace[2pt]
Background familiarity
  & Did the participant feel their background prepared them to answer questions about the unfamiliar paper?
  & \cbvalues{Category}{Not at all, a little bit, very much.} & 70 \\
\addlinespace[2pt]
Question difficulty
  & The participant's overall perception of question difficulty.
  & \cbvalues{Category}{Too easy; Appropriate; Too difficult} & 64 \\
\addlinespace[2pt]
Rubric appropriate
  & The participant thought the grading rubrics were appropriate.
  & \cbvalues{Category}{Disagree, Neither agree nor disagree, Agree} & 61 \\
\addlinespace[2pt]
Deployment
  & Should a system like this be deployed in conferences or journals?
  & \cbvalues{Category}{No, Maybe with Significant Changes, Yes} & 60 \\
\addlinespace[2pt]
Perceived purpose
  & Did the participant guess the intended purpose of the system?
  & \cbvalues{Category}{No, Close Guess, Yes} & 55 \\
\addlinespace[2pt]
Time pressure
  & Did the participant feel pressured for time?
  & \cbvalues{Binary}{Present, Absent} & 55 \\
\addlinespace[2pt]
Overall preference
  & How did the participant like the overall experience with the system?
  & \cbvalues{Category}{Disliked, Neither liked nor disliked, Liked} & 54 \\
\addlinespace[2pt]
Grading agreement
  & The participant agreed with their grades.
  & \cbvalues{Category}{Disagree, Neither agree nor disagree, Agree} & 51 \\
\addlinespace[2pt]
Question depth
  & Whether the question requires applying, extending, critiquing, or reconstructing ideas beyond recalling the text.
  & \cbvalues{Category}{Not tested; Weakly tested; Strongly tested} & 44 \\
\addlinespace[2pt]
Trust change
  & How would the use of our system change the participant's trust in a submission process?
  & \cbvalues{Category}{Decrease Trust, Neither increase nor decrease trust, Increase trust} & 41 \\
\addlinespace[2pt]
Circumvention feasibility
  & How feasible the participant thinks it would be to prepare strategically without genuinely verifying the paper.
  & \cbvalues{Category}{Low; Moderate; High} & 31 \\
\addlinespace[2pt]
Feedback relevance
  & The participant found the feedback provided by the system relevant.
  & \cbvalues{Category}{Disagree, Neither agree nor disagree, Agree} & 28 \\
\addlinespace[2pt]
Contribution coverage
  & Whether important parts of the participant's contribution were tested.
  & \cbvalues{Category}{Important omissions; Partial coverage; Strong coverage} & 27 \\
\addlinespace[2pt]
Grading false positive
  & The participant reports that a weak or incorrect answer received too much credit.
  & \cbvalues{Binary}{Present, Absent} & 7 \\
\addlinespace[2pt]
Grading false negative
  & The participant reports that a correct or reasonable answer received too little credit.
  & \cbvalues{Binary}{Present, Absent} & 6 \\
\bottomrule
\end{tabularx}
\end{table*}

\clearpage
\section{Inductive Codebook}
\label{apx:inductive_codebook}


\begingroup
\small
\setlength{\tabcolsep}{5pt}
\renewcommand{\arraystretch}{1.04}
\setlength{\LTleft}{0pt}
\setlength{\LTright}{0pt}
\setlength{\LTcapwidth}{\linewidth}

\newcommand{\inductiveentry}[3]{%
  \textbf{#1}\par
  {\itshape #2\par}%
  &
  \begin{tabular}[t]{@{}
    >{\RaggedRight\arraybackslash}p{%
      \dimexpr\linewidth-3em-2\tabcolsep\relax}
    >{\raggedleft\arraybackslash}p{3em}
  @{}}
    #3
  \end{tabular}
  \\ \addlinespace[0.55em]
}

\begin{longtable}{@{}
  >{\RaggedRight\arraybackslash}p{0.36\linewidth}
  >{\RaggedRight\arraybackslash}p{%
    \dimexpr0.64\linewidth-2\tabcolsep\relax}
@{}}
\caption{Inductive codes and counts, grouped by category.
Categories are ordered by the sum of their code counts;
codes are ordered by count within each category.}
\Description{A three-page table listing 61 inductive interview codes grouped into 16 categories. Each category includes a short definition, followed by its code names and occurrence counts. Categories address expectations, emotions, preparation, time and content limits, proposed changes to the assessment, deployment roles and concerns, unintended measurements, experience confounds, learning, grading errors, answer sources, field-level validity, and interface design. The most frequent codes include time pressure with 55 occurrences, the system expecting a particular answer with 52, uncertainty about the required level of detail with 44, reflection on performance with 32, and frustration with 26.}

\label{tab:inductive-codes}\\

\toprule
\textbf{Category and description}
& \textbf{Code}\hfill\textbf{Count} \\
\midrule
\endfirsthead

\multicolumn{2}{@{}l@{}}{%
  \small\itshape Table~\ref{tab:inductive-codes} continued.
}\\
\toprule
\textbf{Category and description}
& \textbf{Code}\hfill\textbf{Count} \\
\midrule
\endhead

\midrule
\multicolumn{2}{@{}r@{}}{%
  \footnotesize\itshape Continued on next page.
}\\
\endfoot

\bottomrule
\endlastfoot

\inductiveentry{Clarify expectations}{
  Uncertainty about the expected answer, level of detail, or criteria for an acceptable response.
}{
  System expects particular answer & 52 \\
  Unclear how much detail expected & 44 \\
  Transparency of expected outcome & 13
}

\inductiveentry{Negative emotions}{
  Emotional reactions to assessment and feedback.
}{
  Reflection on performance & 32 \\
  Frustration & 26 \\
  Feels like on trial & 14 \\
  Black mirror & 11 \\
  Breakdown of trust & 2
}

\inductiveentry{How to prepare}{
  Strategies for preparing for the assessment through reading,
  careful note-taking, AI assistance, and critical examination
  of the paper.
}{
  Prepare by skimming & 24 \\
  Prepare with LLM & 20 \\
  Prepare by documentation & 17 \\
  Inspect references & 11 \\
  Find inconsistencies & 4
}

\inductiveentry{Content limit}{
  Constraints on demonstrating understanding by time pressure
  and response-length limits.
}{
  (Deductive: Time pressure) & 55 \\
  Length limit & 13
}

\inductiveentry{Change modality or test structure}{
  Proposed changes to questions, rubrics, interaction,
  and response modalities to improve the assessment
  during deployment.
}{
  Question changes & 20 \\
  Interaction with system & 19 \\
  Rubric changes & 17 \\
  System modality & 6
}

\inductiveentry{Deployment role}{
  Proposed roles for greCAPTCHA results in downstream
  decisions, including who should receive them.
}{
  Desk reject or ban & 18 \\
  Threshold choice & 17 \\
  Show to reviewers & 12 \\
  Just show score to authors & 5 \\
  Opt-in choice & 3 \\
  Show to metareviewers & 2
}

\inductiveentry{Deployment concern}{
  Practical and governance concerns about deployment.
}{
  Deployment overhead & 21 \\
  Requires human oversight & 17 \\
  Need for deployment transparency & 7 \\
  Deployment with many projects & 5 \\
  Deployment with many collaborators & 3 \\
  Seniority effects & 1
}

\inductiveentry{Unintended measurements}{
  Concerns that scores reflect unintended capabilities
  such as test-taking skills, which are beyond the
  participant's control.
}{
  Measures typing ability & 17 \\
  Measures test aptitude & 10 \\
  Measures speed of thinking & 8 \\
  Measures English fluency & 6 \\
  Measures searching ability & 5 \\
  Measures paper age & 3
}

\inductiveentry{Experience confound}{
  Perceived influences of disciplinary background,
  field familiarity, and specific research contributions
  on greCAPTCHA performance.
}{
  Background dependence & 21 \\
  Field dependence & 12 \\
  Contribution dependence & 11
}

\inductiveentry{Learning}{
  Opportunities to use the assessment for learning,
  reflection, and identifying gaps in understanding
  one's own research.
}{
  Use case for learning & 25 \\
  Learn about own paper & 19
}

\inductiveentry{Uncategorized}{
  Additional observations about assessment experiences,
  alternative approaches, and contextual factors
  not captured by other categories.
}{
  Baffled by unfamiliar paper & 13 \\
  Alternative systems for authorship check & 7 \\
  Question order & 6 \\
  Extenuating reason for failure & 4 \\
  Response time & 3 \\
  Communication importance & 2
}

\inductiveentry{Preparation time}{
  Accounts of the amount of time spent preparing
  for the unfamiliar-paper assessment.
}{
  Short preparation & 17 \\
  Long preparation & 9
}

\inductiveentry{False positives and false negatives}{
  Perceived errors in awarding or withholding credit,
  and concerns about incorrectly marking authors
  as non-authors.
}{
  (Deductive: Grading false positive) & 7 \\
  (Deductive: Grading false negative) & 6 \\
  System weeds out wrong people & 5
}

\inductiveentry{Answer source}{
  Resources used or discussed regarding answering
  questions, including background knowledge,
  the manuscript, and cheating.
}{
  Use background & 9 \\
  Cheating & 4 \\
  Use paper & 2
}

\inductiveentry{Perceived validity: field}{
  Views on whether the assessment provides valid
  evidence of understanding the manuscript's
  research field.
}{
  Agree & 7 \\
  Disagree & 4 \\
  Neither agree nor disagree & 0
}

\inductiveentry{UI design}{
  Experiences of interface clarity and how it
  supports or obstructs interaction with
  the assessment.
}{
  UI unclear & 9 \\
  UI clear & 2
}

\end{longtable}
\endgroup

\end{document}